\PassOptionsToPackage{dvipsnames}{xcolor}
\documentclass[twocolumn]{aastex631}

\usepackage{graphicx}	
\usepackage{amsmath}	
\usepackage{amssymb}	
\usepackage{mathtools}
\usepackage[left,mathlines]{lineno}

\usepackage{ulem}
\usepackage{soul}
\usepackage{comment}
\usepackage{bm}

\usepackage{macros}

\begin{document}
\bibliographystyle{aasjournal}


\shorttitle{Constraining the Stellar Mass Halo Mass Relation}
\shortauthors{}

\title{Constraining the Low Mass End of the Stellar Halo Mass Relation with Surveys of Satellite Galaxies}

\correspondingauthor{J. Sebastian Monzon}
\email{s.monzon@yale.edu}

\author[0000-0002-9986-4604]{J. Sebastian Monzon}
\affiliation{Department of Astronomy, Yale University, PO. Box 208101, New Haven, CT 06520, USA}

\author[0000-0003-3236-2068]{Frank C. van den Bosch}
\affiliation{Department of Astronomy, Yale University, PO. Box 208101, New Haven, CT 06520, USA}

\author[0000-0001-8073-4554]{Kaustav Mitra}
\affiliation{Department of Astronomy, Yale University, PO. Box 208101, New Haven, CT 06520, USA}


\begin{abstract}
The abundance of satellite galaxies is set by the hierarchical assembly of their host halo. We leverage this to investigate the low mass end ($M_\rmH < 10^{11} \Msun$) of the Stellar-to-Halo Mass Relation (SHMR), which is key to constraining theories of galaxy formation and cosmology. We argue that recent analyses of satellite galaxies in the Local Group environment have not adequately modelled the dominant source of scatter in satellite stellar mass functions: the variance in accretion histories for a fixed host halo mass. We present a novel inference framework that not only properly accounts for this halo-to-halo variance but also naturally identifies the amount of host halo mass mixing, which is generally unknown. Specifically, we use the semi-analytical SatGen model to construct mock satellite galaxy populations consistent with the third data release of the Satellites Around Galactic Analogs (SAGA) survey. We demonstrate that even under the most idealized circumstances, the halo-to-halo variance makes it virtually impossible to put any meaningful constraints on the scatter in the SHMR. Even a satellite galaxy survey made up 100 hosts can at best only place an upper limit of $\sim 0.5$dex on the scatter (at the 95\% confidence level). This is because the large variance in halo assembly histories dominates over the scatter in the SHMR. This problem can be overcome by increasing the sample size of the survey by an order of magnitude ($\sim 1000$ host galaxies), something that should be fairly straightforward with forthcoming spectroscopic surveys.
\end{abstract}  


\section{Introduction}
\label{sec:intro}

The advent of large galaxy (redshift) surveys has allowed for tight constraints on the galaxy-halo connection; the statistical relationship between galaxies and the dark matter halos they inhabit \citep[see][for a detailed review]{Wechsler.Tinker.18}. Using a variety of different summary statistics, including galaxy clustering \citep[e.g.][]{Yang.etal.03, Tinker.etal.12, Zehavi.etal.11}, galaxy-galaxy lensing \citep[e.g.,][]{Mandelbaum.etal.06, Cacciato.etal.13, Leauthaud.etal.12}, galaxy group catalogues \citep[e.g.,][]{Weinmann.etal.06, Yang.etal.07, Tinker.etal.21}, and satellite kinematics \citep[e.g.,][]{vdBosch.etal.04, More.etal.11, Lange.etal.19b,  Mitra.etal.24}, numerous studies have established that central galaxies follow a fairly narrow stellar-to-halo mass relation (hereafter SHMR).

The SHMR, $\mstar(M_\rmH)$, is well characterized by a broken power-law relation with a break near $M_\rmH = 10^{12} \Msun$ \citep[e.g.,][]{Yang.etal.03, Vale.Ostriker.06, Guo.etal.10, Moster.etal.10}. At the massive end ($M_\rmH > 10^{12}\Msun$), the stellar mass of central galaxies increases slowly with increasing halo mass, with $\rmd\log \mstar/\rmd\log M_\rmH \sim 0.3$ \citep[e.g.,][]{Behroozi.etal.10, More.etal.11, Yang.etal.12, Wechsler.Tinker.18}. In addition, the scatter in the SHMR, $\sigma$, which is typically characterized as the log-normal distribution in stellar mass at given halo mass, is inferred to be fairly small with most studies finding the scatter to be between 0.15 and 0.20dex \citep[e.g.,][]{Moster.etal.10, More.etal.11, Leauthaud.etal.12, Cacciato.etal.13, Zu.Mandelbaum.15, Porras-Valverde.etal.2023, Mitra.etal.24}. Below a halo mass scale of $M_\rmH \sim 10^{11} \Msun$, though, both the slope, $\alpha \equiv \rmd\log \mstar/\rmd\log M_\rmH$, and scatter of the SHMR are extremely poorly constrained \citep[e.g.,][]{Brook.etal.14, Wechsler.Tinker.18, Jethwa.etal.18, Allen.etal.19, Sales.etal.22}. This is because galaxies that are hosted by halos with $M_\rmH \lta 10^{11} \Msun$ are too faint to be seen out to a significant distance and therefore cannot be a representative volume sample of the Universe. 

In principle, this shortcoming can be overcome if one accounts for the specific aspects of the volume probed by the data. A good example of this are attempts to constrain the low-mass end of the SHMR using data of satellite galaxies in the Milky Way (hereafter MW) halo or Local Group (hereafter LG). Typically, these studies use simulations or semi-analytical techniques to predict the dark matter mass distribution (in particular the abundance of halos and subhalos) in environments that resemble the MW/LG. Two different, but related, techniques are then used to constrain the SHMR: (i) forward modeling, where a model for the SHMR is combined with the masses of the subhalos at accretion, to predict the (present-day) stellar masses of the surviving satellite galaxies \citep[e.g.,][this study]{Garrison-Kimmel.etal.17b, Jethwa.etal.18, Santos.etal.22, Danieli.etal.23}, or (ii) abundance matching, in which the SHMR is established empirically by matching an observed satellite galaxy, rank-ordered by stellar mass, to a subhalo, rank-ordered by halo mass at accretion or by peak circular velocity\footnote{Typically some scatter in the rank-order matching is included, which translates to scatter in the SHMR, which would otherwise be completely deterministic.} \citep[e.g.,][]{Brook.etal.14, Read.etal.17, Nadler.etal.19a, Nadler.etal.24}.

These methods have to overcome several challenges. First of all, despite the dramatic increase in the number of satellite/dwarf galaxies detected in the MW/LG environment, largely driven by the Sloan Digital Sky Survey (SDSS) and Dark Energy Survey (DES) \citep[see][and references therein]{Simon.19, DrlicaWagner.etal.15}, we still do not have a complete census. Consequently, incompleteness corrections have to be folded in which introduces significant uncertainties. Secondly, it is well established that the abundance of substructure depends strongly on both the host halo's mass and assembly history \citep[e.g.,][]{Gao.etal.04, vdBosch.etal.05a, Giocoli.etal.10, Jiang.vdBosch.17}. Hence one needs to properly account for both the non-negligible uncertainty in the (total) mass of the MW or LG \citep[see][and references therein]{BlandHawthorn.Gerhard.16} as well as the variance in accretion histories at a fixed host halo mas (what we refer to as ``halo-to-halo" variance). Thirdly, the abundance of (sub)halos in a MW halo or LG environment is typically taken from $N$-body simulations which are known to be affected by artificial subhalo disruption \citep[e.g.,][]{vdBosch.etal.18a, vdBosch.etal.18b, Errani.Penarrubia.20} and/or the presence of a central disk galaxy \citep[][]{DOnghia.etal.10, Garrison-Kimmel.etal.17, Green.vdBosch.22}. Finally, both the forward modelling method and the abundance matching technique have their own shortcomings. The former relies on making an assumption about the functional form of the SHMR, $\mstar(M_\rmH)$, at the low mass end. Virtually every study to date has assumed a simple power-law relation with some log-normal scatter, and without any redshift dependence. However, there is no a-priori reason why the SHMR can't be significantly more complicated. Abundance matching, in turn, relies on the notion that the rank-order in stellar mass is tightly correlated with the mass or circular velocity of its host halo. While it has been demonstrated to work well at the massive end \citep[e.g.,][]{Conroy.Wechsler.09, Guo.etal.10, Reddick.etal.13}, where the scatter in the SHMR is small, there is no guarantee that it yields reliable results at the low-mass end where the scatter might well be significantly larger \citep[e.g.,][]{Fitts.etal.18, Munshi.etal.21, Engler.etal.21b}. This is only complicated by the fact that at the low mass end the occupation fraction of dark matter halos is expected to drop below unity due to reionization and the hydrogen cooling limit \citep[e.g.,][]{Efstathiou.92, Thoul.Weinberg.96, Okamoto.etal.08, Bovill.Ricotti.09}. In principle this can be taken into account \citep[see e.g.,][]{Jethwa.etal.18, Nadler.etal.24}, but unfortunately there is no consensus yet as to how exactly the occupation fraction scales with halo mass \citep[e.g.,][]{Sawala.etal.16, Fitts.etal.18, Wheeler.etal.19, Katz.etal.20}, adding additional degrees of freedom to the modeling.

Given all these issues, it should come as no surprise that, to date, the various attempts to constrain the low mass end of the SHMR have reached little to no consensus. Assuming that the SHMR\footnote{From here on out, the ``SHMR" refers to the low-mass end of relation ($M_\rmH \lesssim 10^{11} \Msun$) unless otherwise stated.} follows a power-law relation of the form $\mstar \propto M_\rmH^{\alpha}$, different authors report wildly different constraints on the value of $\alpha$ ranging all the way from $\sim 1.5$ \citep[e.g.][]{Behroozi.etal.13c, Read.etal.17} to values larger than $3.0$ \citep[e.g.,][]{Yang.etal.12, Brook.etal.14} and everything in between \citep[see Fig.~15 of][]{Jethwa.etal.18}. In addition, little is known about the scatter in the SHMR at the low mass end, other than that its inferred value is found to be degenerate with the inferred, or assumed, value of $\alpha$ \citep[][]{Garrison-Kimmel.etal.17b}. In fact, an important outstanding question of relevance for our understanding of galaxy formation, is whether the scatter continues to remain small (i.e., comparable to the $0.2$dex scatter at the massive end), or whether the scatter significantly increases towards lower mass halos. 

Unfortunately, little guidance is provided by hydro-dynamical simulations, which reveal a similar lack of consensus when it comes to predicting the SHMR.  Although most simulations predict that the SHMR roughly continues as a power-law down to the smallest halos they can resolve ($\sim 10^8 - 10^9 \Msun$), the predicted slope, scatter and normalization differ substantially from simulation to simulation \citep[see][and references therein]{Sales.etal.22}. In particular, whereas most simulations predict relatively little scatter in the SHMR across the mass range probed \citep[but see][]{Munshi.etal.13, Fitts.etal.18}, the simulation-to-simulation scatter is typically much larger. In fact, at a halo mass of $\sim 10^9 \Msun$, the {\it average} stellar mass predicted by different simulations differs by more than two orders of magnitude \citep[][]{Sales.etal.22, Garrison-Kimmel.etal.17b}. In addition, in order to achieve sufficient resolution to resolve dwarf galaxies, these studies typically use simulations that zoom-in on individual halos. Since these simulations are CPU-intensive, building a sufficiently large sample to adequately capture halo-to-halo variance is challenging.

Despite the current lack of consensus, there are good reasons to be optimistic. In the near future, new and ongoing surveys such as Pan-STARRS \citep[PS1][]{Chambers.etal.16}, the Dark Energy Survey \citep[DES][]{Abbott.etal.18} and the Dark Energy Spectroscopic Instrument Survey \citep[DESI][]{Desi.etal.22}, as well as new observational facilities such as the Vera C. Rubin Observatory \citep[][]{Ivezic.etal.19} and the Nancy Grace Roman Space Telescope \cite[][]{Spergel.etal.15}, will dramatically improve our census of nearby dwarf/satellite galaxies. In addition, as detailed in Section~\ref{sec:SAGA} below, several surveys are underway that are specifically dedicated to measuring the satellite stellar mass functions (hereafter SSMF) around a significant sample of host galaxies.  

In this paper we investigate what kind of constraints on the SHMR one may hope to infer when using stellar mass functions from surveys of satellite galaxies around MW-like hosts. This is somewhat similar, but complementary to, two recent studies; \citet{Nadler.etal.24} and \citet[][hereafter D23]{Danieli.etal.23}. Whereas the former forecasts constraints from a complete census of MW satellites (i.e., down to the cutoff halo mass below which galaxy formation is suppressed), the latter constrains the SHMR using a census of satellite galaxies within $\sim 12 \Mpc$ of the MW. Our analysis differs from these studies in that (i) we use much larger samples of halos, thereby properly accounting for halo-to-halo variance, (ii) we adopt a different, more detailed, statistical analysis of the data, and (iii) we consider more flexible functional forms for the SHMR. We show that it is extremely challenging to put any meaningful constraints on the scatter in the SHMR, even with a perfect (i.e., no completeness corrections, no interlopers, no observational errors on stellar masses, etc.) version of the current state-of-the-art in satellite galaxy surveys. This is mainly because even these surveys are not yet large enough to beat down the dominant source of variance in the data; halo-to-halo variance.

This paper is organized as follows: We start in Section~\ref{sec:SAGA} with a brief overview of existing surveys of satellite galaxies around MW-like hosts. Section~\ref{sec:method} describes the construction of mock data and the Bayesian inference framework we developed. This includes a brief overview of the \SatGen model that we use to construct a large sample of subhalo populations of MW-like host halos, a description of our flexible SHMR model, and the statistics we use to characterize the satellite galaxy populations. Section~\ref{sec:res} presents our main parameter recovery results, while Section~\ref{sec:mass_mixing} examines the impact of incorrect priors regarding the distribution of host halo masses. We finish with a discussion of how and why our results differ from previous work (Section~\ref{sec:discussion}) and a summary of our main conclusions (Section~\ref{sec:conclusions}). Throughout, we adopt a flat $\Lambda$CDM cosmology with a present-day matter density $\Omega_\rmm = 0.3$, a baryon density $\Omega_\rmb = 0.0465$, power spectrum normalization $\sigma_8 = 0.8$,  spectral index $n_\rms = 1.0$, and a Hubble parameter  $h = (H_0/100\kmsmpc) = 0.7$, roughly in agreement with the Planck18 constraints \citep[][]{Planck.18}.

\section{Surveys of Satellite Galaxies}
\label{sec:SAGA}

In recent years our census of satellite galaxies around Milky Way-like hosts has increased drastically, mainly due to two dedicated surveys, the Exploration of Local VolumE Satellites survey (ELVES) and the Satellites Around Galactic Analogs (SAGA) Survey, that yield statistically significant samples of bright and classical dwarf satellite galaxies with stellar masses in the range $10^{6} \Msun \lta \mstar \lta 10^9 \Msun$. These surveys serve as a test-bed for hydro-dynamical simulations of (dwarf) galaxy formation, give valuable insights into the quenching of satellite galaxies \citep[][]{Wang.etal.24}, and provides crucial information for constraining the SHMR. 

The ELVES Survey \citep[][]{Carlsten.etal.20, Carlsten.etal.22, Danieli.etal.23} uses deep imaging data to survey the satellite populations of 30 massive host galaxies (with a $K_s$-band magnitude $M_{K_s} < -22.1$ mag) out to a distance of $12 \Mpc$. To date, this has resulted in a few hundred satellite galaxies out to a projected distance of $300\kpc$ from their respective hosts. D23 used the ELVES data to place tight constraints on both the slope and scatter of the low mass end of the SHMR. However, as discussed in detail in Section~\ref{sec:comp}, their analysis has several shortcomings that warrant a more detailed investigation of how best to constrain the SHMR using satellite galaxies.

The SAGA Survey \citep[][]{Geha.etal.17, Mao.etal.21, Geha.etal.24, Mao.etal.24} is a recently completed {\it spectroscopic} survey that has characterized the satellite galaxy populations of 101 MW analogs at $z \sim 0.01$ down to an absolute magnitude of $M_r = -12.3$, roughly equivalent to that of Leo I in the LG. The SAGA survey defines MW analogs as isolated galaxies with an absolute $K$-band magnitude $-23 > M_K > -24.6$ and with no other bright ($K < K_{\rm host} - 1.6$) galaxies within the projected radius of $300\kpc$ of the host (roughly equivalent to the host's virial radius). SAGA satellites are defined to lie within $300\kpc$ and $\pm 275\kms$ from their host galaxies, in projection, and are required to have an (extinction-corrected) $r$-band magnitude $M_r < -12.3$, which roughly corresponds to a stellar mass of $10^{6.5} \Msun$ \citep[][]{Font.etal.20,Karunakaran.etal.21}. The SAGA collaboration recently published their final catalogue \citep[][]{Mao.etal.24} which contains 378 satellite galaxies around 101 MW-mass systems. 

To target the low-mass end of the SHMR, we choose to focus on satellite galaxies around MW-like hosts. Given that the SAGA survey has the largest sample of spectroscopically confirmed satellites, and is the most homogeneous dataset of MW-like host galaxies, we use it as our fiducial template. Throughout, we therefore focus on ``SAGA-like'' surveys, which are surveys of satellite galaxies around 100 MW-like hosts that have a similar depth as the SAGA survey. We emphasize, though, that our results are generic and apply equally well to ELVES or any other (future) survey of satellite galaxies. In the remainder of this paper, we construct and analyze mock data, which we then subject to a detailed statistical analysis in order to examine what we may hope to learn about the low-mass end of the SHMR. In doing so, we take the most optimistic approach. In particular, we ignore potential incompleteness issues due to surface brightness limits, incomplete coverage, unsuccessful spectroscopic measurements, etc. We also ignore complications due to the contamination with interlopers and observational errors (for example in the derived stellar masses). We are aware that these assumptions are unrealistic. Yet, as we will demonstrate, even under these optimal conditions, and with perfect knowledge of the halo masses of the host galaxies, it is extremely challenging to put meaningful constraints on the low mass end of the SHMR. The main purpose of this paper is to emphasize the importance of properly accounting for halo-to-halo variance, the non-negligible uncertainties in the host halo masses, and to advocate a new statistical method to analyze the data.

\section{Methodology}
\label{sec:method}

We construct and analyze mock populations of satellite galaxies as follows. Using the semi-analytical code \SatGen, we create detailed merger trees of MW-like dark matter host halos. As described in Section~\ref{subsec:SatGen}, these merger trees yield the masses, $\macc$, and redshifts, $\zacc$, of subhalos at accretion. We then assign a stellar mass to each of these subhalos using a model for the SHMR of the form $\mstar = \mstar(\macc, \zacc)$ described in Section~\ref{subsec:SHMR}.  Next, we make the highly idealized assumption that the evolution of a galaxy, and thus its stellar mass, is frozen the moment it becomes a satellite galaxy. Hence, the present-day stellar mass of a satellite galaxy is simply assumed to be equal to that at accretion. Although this is a standard assumption made (implicitly) in abundance matching and (explicitly) in the forward modelling of satellite galaxy populations, it is important to be aware that in reality the stellar mass of a satellite can continue to undergo changes (see Section~\ref{subsec:mstar_evo} for a brief discussion of the potential consequences of this oversimplification). 

The scheme outlined above allows us to translate a set of MW-like host halo merger trees into a mock SAGA-like survey of satellite stellar masses. Section~\ref{subsec:data} describes the novel summary statistics that we use to characterize this data, and Section~\ref{sec:stats} discusses the Bayesian framework used to infer constraints on the SHMR from the mock data. 

\subsection{The SatGen Model}
\label{subsec:SatGen}

\begin{figure*}
    \centering
    \includegraphics{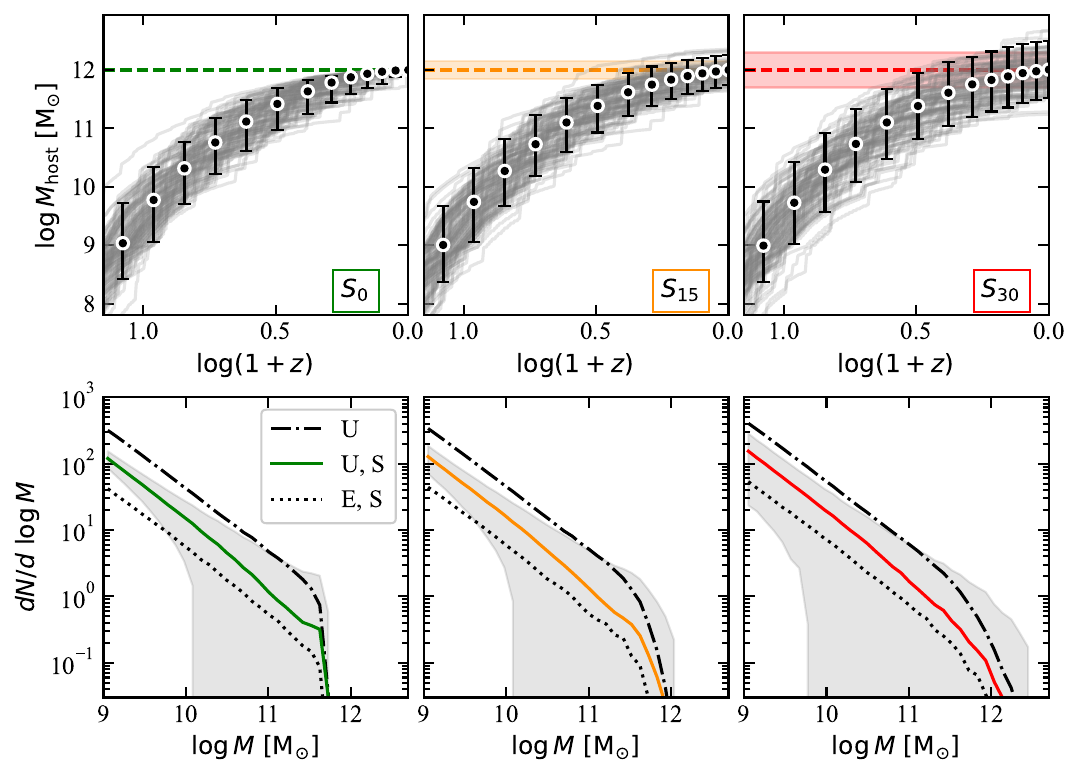}
    \caption{The three SatGen merger tree samples used in this paper: $S_0$ (green), $S_{15}$ (orange) and $S_{30}$ (red). \textbf{Top panels}: 100 randomly selected mass accretion histories (MAHs) from the 10,100 realizations in each sample, depicting the mass of the host halo's main progenitor as a function of redshift. The black points show the median main progenitor mass at different epochs, with the error bars indicating the $5-95$ percentile range across the full sample. The colored horizontal bands show the $16-84$ percentile range of the distribution of the $z=0$ host halo masses. \textbf{Bottom panels}: The average subhalo mass functions (SHMFs) and associated variance across all merger tree realizations in each sample. The solid colored lines are the ``unevolved, surviving" (U,S) SHMFs, which indicate the abundance of subhalos surviving to $z=0$, as a function of their mass at accretion, $\macc$. The grey contours show the corresponding $16-84$ percentile ranges across the $10,100$ hosts. For comparison, the dash-dotted black lines show the ``unevolved" (U) SHMFs, indicating the abundance of all subhalos ever accreted onto the host as a function of $\macc$, while the dotted, black lines are the ``evolved, surviving" (E, S) SHMFs, indicating the abundance of surviving subhalos as a function of their present-day ($z=0$) mass. Together, these three SHMFs illustrate the effects of disruption (from dash-dotted to solid) and tidal stripping (from solid to dotted). }
    \label{fig:shmfs}
\end{figure*}

Rather than using numerical $N$-body simulations, we construct merger histories and subhalo populations for a large sample of MW-like host halos using \SatGen \citep[][]{Jiang.etal.21, Green.etal.21}, a state-of-the-art, semi-analytical model devised to generate statistical samples of satellite galaxy populations for hosts of a given mass in a given cosmology. It is orders of magnitude faster than using zoom-in cosmological simulations, gives full freedom in picking the exact masses of the host halos, and most importantly, is free from numerical issues related to artificial disruption and unreliable subhalo finders. 

Briefly, \SatGen constructs merger trees for host halos of a given mass and a given cosmology using the method of \citet{Parkinson.etal.08}. These yield the accretion masses, $\macc$, and redshifts, $\zacc$, of the subhalos that have been accreted by the halo's main progenitor over its entire history, and are in excellent agreement, in a statistical sense, with merger trees extracted from $N$-body simulations \citep[][]{Jiang.vdBosch.14, vdBosch.etal.14}. For each subhalo, \SatGen randomly draws the orbital energy and orbital angular momentum using the distributions of \cite{Li.etal.20}. Given these initial conditions, \SatGen integrates the orbit in the host halo (assumed to be spherical) while accounting for dynamical friction, tidal stripping and tidal heating using analytical recipes. All of these components have been carefully calibrated against a large suite of several thousand, high-resolution, idealized simulations \citep[][]{Ogiya.etal.19, Jiang.vdBosch.16, Jiang.etal.21, Green.etal.21}. 

Although \SatGen has great flexibility for modelling satellite galaxies \citep{Folsom.etal.2023}, including a treatment of the stellar and gaseous components and the impact of baryonic feedback on the (sub)halo density profiles \citep[see][for details]{Jiang.etal.21}, we do not make use of this functionality. Rather, in this study we only use the dark-matter only version of \SatGen to construct the merger trees and to model the tidal evolution of the subhalos. Satellite galaxies will be pasted on to the subhalos separately as described in Section~\ref{subsec:SHMR} below.

We use \SatGen to generate three samples of $10,100$ merger trees each. Halo masses are drawn from a log-normal distribution centered on $M_\rmH = 10^{12} \Msun$ and with a scatter of $\sigma_M$dex. For the first sample, $S_0$, we adopt $\sigma_M=0$ so that all host halos have exactly the same halo mass of $10^{12} \Msun$. The subsequent samples, $S_{15}$ and $S_{30}$, are constructed with $\sigma_M = 0.15$ and $0.30$dex, respectively. Each of these samples allows us to construct 101 surveys, each consisting of 100 host halos. One of these surveys will serve as the mock data, while the others will be used in the analysis (see Section~\ref{subsec:mock} below).

Each merger tree traces the evolution of each of its branches back to a `leaf-mass' of $\mleaf = 10^{9}\Msun$, which specifies our effective mass-resolution. This is more than adequate to model satellite galaxies in a SAGA-like survey, which are limited to satellites with stellar mass in excess $\mlim \sim 10^{6.5} \Msun$ (see Section \ref{sec:SAGA}). Each subhalo is evolved in the host halo's main progenitor from its time of accretion to the present time, accounting for dynamical friction and tidal mass loss as detailed in \citet{Green.vdBosch.19} and \citet{Green.etal.21}.

Throughout, we assume that a subhalo (and its associated satellite galaxy) are disrupted whenever the instantaneous subhalo mass drops below $m_{\rm dis} \equiv f_{\rm dis} \macc$. In practice, we account for this tidal disruption by only populating subhalos with satellite galaxies if the subhalo's present-day mass exceeds $m_{\rm dis}$. We adopt $f_{\rm dis} = 10^{-4}$ as our fiducial value, which implies that subhalos disrupt once they have lost more than 99.99 \% of their initial accretion mass due to tidal stripping. We have verified that setting $f_{\rm dis} = 10^{-3}$ instead has no significant impact on any of our results.

The upper panels of Fig.~\ref{fig:shmfs} show the mass accretion histories (MAHs) for a random selection of 100 host halos in samples $S_0$, $S_{15}$ and $S_{30}$, as indicated.  Note how the scatter in main progenitor masses is comparable among all three samples at high redshift. At lower redshifts, though, the scatter is clearly larger for the samples with the larger variance in host halo masses. As we will see, this scatter in host halo masses induces strong covariance in the satellite populations of host galaxies in a SAGA-like survey. The lower panels of Fig.~\ref{fig:shmfs} show the corresponding average subhalo mass functions (SHMFs) obtained from the full samples of $10,100$ halo merger trees. The solid, colored curves show the ``unevolved-surviving" SHMFs, i.e., the abundance of subhalos that survive to the present-day as a function of their mass at accretion, $\macc$. These are the mass functions that we use to construct our satellite populations. The shaded contours mark the $16-84$ percentile ranges of the unevolved-surviving SHMFs, highlighting the enormous variance in subhalo statistics, especially at the massive end \citep[see also][]{Jiang.vdBosch.17}. Note also that the variance becomes appreciably larger with increasing scatter in the masses of the host halos (i.e., from left to right). Importantly, while the $S_0$ sample is only affected by ``halo-to-halo" variance, the $S_{15}$ and $S_{30}$ samples show the combined effects of halo-to-halo variance plus host halo mass mixing. We refer to this combined effect as ``host-to-host" variance.

\subsection{The Stellar Mass-Halo Mass Relation}
\label{subsec:SHMR}

As discussed in Section~\ref{sec:intro}, it is common practice to assume that the low-mass end of the SHMR can be characterized by a simple power-law relation of the form $\mstar \propto M_\rmH^{\alpha}$, often with a fixed amount of log-normal scatter in stellar mass at fixed halo mass. We define this simple power law as:
\begin{align}\label{SHMR_PL}
 \log[\mstar] & = \log[\mstaranchor] + \alpha \log[M_{12}] + \calG(\sigma)\,. 
\end{align}
Here $M_{12} \equiv \macc - 10^{12} \Msun$, $\alpha$ is the power-law slope, and $\calG(\sigma)$ is a normally distributed deviate with zero mean and variance $\sigma^2$. In order to anchor the SHMR to existing constraints at the high-mass end \citep[][]{Behroozi.etal.13c, RodriguezPuebla.etal.17}, we set $\mstaranchor = 10^{10.5}\Msun$ throughout. Hence, we ignore any potential uncertainty in this normalization, something that an analysis of real data should probably take into account\footnote{We have verified that allowing for $0.1-0.3$dex freedom in $\mstaranchor$ slightly weakens the constraints on $\alpha$, but has virtually no impact on the constraints on $\sigma$.}.

With $\mstaranchor$ fixed, our model for the SHMR is characterized by only two free parameters; $\btheta = (\alpha, \sigma)$. As our fiducial model, we use $\thetafid=(2.0,0.2)$. In Section~\ref{subsec:extra_freedom} we explore a more generic form of the SHMR that allows for curvature in the relation, redshift evolution, and a mass dependant scatter.

\subsection{Mock Survey Assumptions}
\label{subsec:mock}

Motivated by the SAGA survey's recent data release, we define our mock surveys as made up of $\Nhost=100$ host systems (100 different merger tree realizations). Throughout we assume that each satellite galaxy above a given stellar mass threshold, $\mlim$, makes it into the mock survey (i.e., our SAGA-like surveys are assumed to be 100\% complete for satellites with $\mstar > \mlim$). In general agreement with SAGA's effective magnitude limit (see Section~\ref{sec:SAGA}), we take $\mlim = 10^{6.5} \Msun$ as our fiducial value. Later, in Section~\ref{subsec:survey_limits}, we investigate how varying $\mlim$ impacts our inference. We also assume that the survey contains no interlopers, and that the stellar masses in the survey have zero error. While we recognize that these assumptions are unrealistically optimistic, the main goal of this paper is to understand what kind of constraints on the SHMR we can expect from satellite surveys around MW-like hosts given the effects of halo-to-halo variance. By  operating under these idealized circumstances, we report what should be interpreted as the ``best-case" scenario. Any subsequent analyses that properly account for these observational sources of uncertainty should yield weaker constraints.
\begin{figure*}    \centering
    \includegraphics{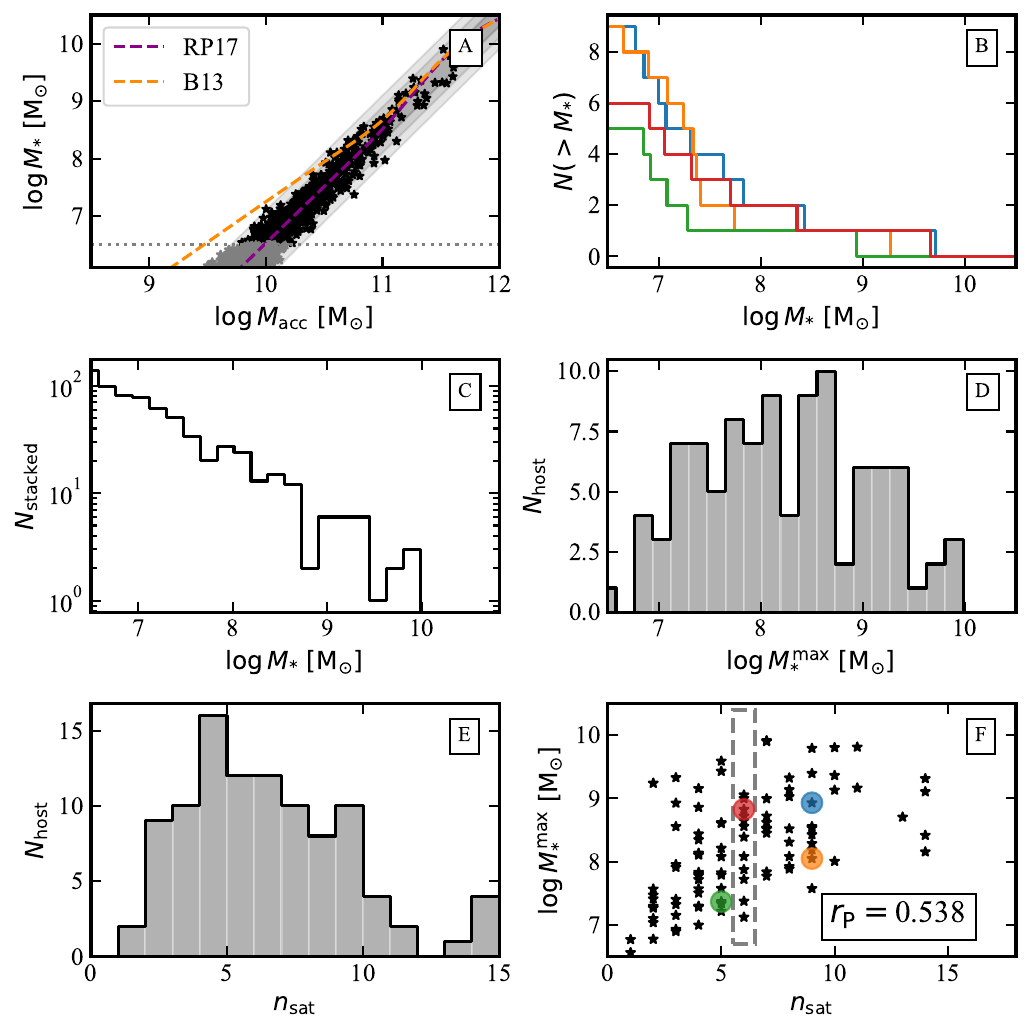}
    \caption{Illustration of various statistics of a mock SAGA-like survey, constructed using the fiducial SHMR model and 100 host halos from sample $S_{15}$. {\bf Panel [A]:} stellar mass as a function of subhalo mass at accretion. The grey shaded regions mark the 68, 95 and 99 percent probability intervals, while the asterisks indicate stellar masses drawn. Note that we only consider satellites with a stellar mass above the mass limit of $10^{6.5} \Msun$ (horizontal dotted line) to be part of the mock data. For comparison, the dashed lines show the SHMR of \citet{RodriguezPuebla.etal.17} and \citet{Behroozi.etal.13c}, as indicated. {\bf Panel [B]:} Cumulative SSMFs for a random subset of 4 host galaxies. {\bf Panel [C]:} The combined or {\it stacked} SSMF for all 100 host galaxies in the mock survey. {\bf Panel [D]:} Histogram showing the distribution of $\mstarmax$ (stellar masses of most massive satellites) for the 100 hosts. {\bf Panel [E]:} Same as panel [D] but for the ``observed" number of satellites, $\nsat$ (i.e., those with $\mstar \geq \mlim$). {\bf Panel [F]:} Scatter plot of $\mstarmax$ versus $\nsat$.  Note that these two quantities are strongly correlated (in this particular sample, Pearson's rank-order correlation coefficient $r_\rmP = 0.538$, as indicated). The colored circles depict the host systems for which the cumulative SSMF is shown, in matching color, in Panel [B]. The dashed rectangle highlights the systems with $\nsat=6$. See text for discussion.}
    \label{fig:summary_stats}
\end{figure*}

\subsection{Statistical Description of the Data}
\label{subsec:data}

There are various summary statistics that can be used to describe data from satellite surveys such as SAGA and ELVES. In this analysis we focus solely on the stellar masses of the satellite galaxies. Additional data such as sizes, colors, or star formation rates are not considered. 

Virtually all attempts to date to constrain the SHMR using data on satellite galaxies have done so by fitting models to the satellite stellar mass function (SSMF), $\rmd N_{\rm sat}/\rmd M_{\ast}$, which simply counts the number of satellite galaxies in a given stellar mass bin, or to the cumulative version thereof, given by
\begin{equation}
N_{\rm sat}(>\mstar) = \int_{\mstar}^{\infty} \frac{\rmd N_{\rm sat}}{\rmd \mstar'} \, \rmd \mstar'\,.
\end{equation}
In their analysis of ELVES data, D23 fit the stacked SSMF, obtained by combining all of their 27 host systems. They then define the likelihood of their data using the formalism of \cite{Nadler.etal.19a, Nadler.etal.20}, which assumes that the satellite counts in different stellar mass bins follow independent Poisson distributions (both in the data and in the model realizations), with an unknown rate parameter which is marginalized over. This implies that their likelihood for the SSMF, described by the data vector $\bD$, given a model vector $\btheta$, can be written as
\begin{equation}\label{calLelves}
    \calL_{\rm SSMF}(\bD| \theta) = \prod_{i=1}^{N_{\rm bin}} P(n_{{\rm obs},i} | \hat{n}_{i,1},..., \hat{n}_{i,\hat{N}}),
\end{equation}
where
\begin{align}
&P(n_{{\rm obs},i} | \hat{n}_{i,1},..., \hat{n}_{i,\hat{N}}) = \left(\frac{\hat{N}+1}{\hat{N}} \right)^{-(\hat{n}_{i,1}+...+\hat{n}_{i,\hat{N}}+1)} \nonumber \\
& \times (\hat{N}+1)^{-n_{{\rm obs},i}} \frac{(\hat{n}_{i,1}+...+\hat{n}_{i,\hat{N}}+n_{{\rm obs},i})!}{n_{{\rm obs},i}! \, (\hat{n}_{i,1}+...+\hat{n}_{i,\hat{N}})!}
\end{align}
Here $n_{{\rm obs},i}$ is the observed number of satellites in stellar mass bin $i$ ($i=1,...,N_{\rm bin}$), and $\hat{n}_{i,j}$ is the number of satellites in the same stellar mass bin in model realization $j$, where $j=1,...,\hat{N}$ \citep[see][for a derivation]{Nadler.etal.19a}. This approach, although powerful, has two shortcomings. First of all, neighboring bins of the SSMF are not independent, but are typically correlated \citep[][]{Jethwa.etal.18}. This correlation largely stems from the fact that more (less) massive hosts have more (fewer) subhalos in all mass bins. In other words, host halo mass mixing introduces covariance in neighboring mass bins. Ignoring this covariance will cause one to infer posteriors that are artificially too tight (see Section~\ref{sec:mass_mixing}). Secondly, and more importantly, by using the {\it stacked} SSMF one loses valuable information regarding the variance in satellite populations from one host to another. We therefore propose another method to characterize the data. We find that two observable quantities carry most of the information about each host system: the total number of satellite above the mass limit of the survey, $\nsat$, and the stellar mass of its most massive satellite, $\mstarmax$. Crucially, we also account for the covariance between these quantities, which holds important information. 

To illustrate this, Fig.~\ref{fig:summary_stats} shows several statistics of a mock SAGA survey that we constructed by populating 100 merger trees taken from sample $S_{15}$ using our fiducial, power-law SHMR model $\thetafid$. Panel (A) plots the satellite stellar masses as a function of their halo mass at accretion, depicting the SHMR used to construct the mock data. Panel (B) plots the cumulative SSMFs for a subset of four host systems, while panel (C) shows the stacked SSMF; all 100 host systems combined. The remaining panels show the summary statistics that we use in this paper. Panel (D) shows the frequency distribution of the (binned) stellar masses of the most massive satellites, $\mstarmax$, panel (E) the frequency distribution of the numbers of satellites around individual hosts, $n_{\rm sat}$, and finally panel (F) shows that $\mstarmax$ and $n_{\rm sat}$ are correlated with, in this case, a Pearson rank-order correlation coefficient $r_\rmP = 0.538$. For context, $r_\rmP \approx 0.74$ in the real SAGA survey (Y. Mao, priv. communication).
\begin{figure}
    \centering
    \includegraphics{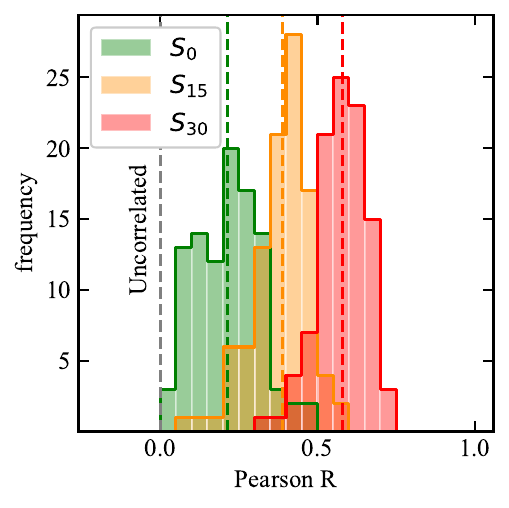}
    \caption{Distributions of the values of Pearson's rank-order correlation coefficient, $r_\rmP$, expressing the correlation between $n_{\rm sat}$ and $\mstarmax$, for 100 SAGA-like surveys constructed using samples $S_0$, $S_{15}$ and $S_{30}$, as indicated. Dotted vertical lines indicate the corresponding medians, while the grey, dashed line marks $r_\rmP=0$, corresponding to no correlation between $n_{\rm sat}$ and $\mstarmax$. See text for discussion.}
    \label{fig:correlation}
\end{figure}

To further highlight the significance of the correlation between $\nsat$ and $\mstarmax$, Fig.~\ref{fig:correlation} shows the distributions of $r_\rmP$ obtained for 100 independent SAGA-like surveys again constructed using our fiducial SHMR. Note how the average value of $r_\rmP$ increases with the amount of host halo mass mixing in the sample (see also Appendix~\ref{app:mah}). This can be understood as follows. More massive halos, on average, host more satellites above a given stellar mass limit. Similarly, more massive halos, on average, accrete more massive subhalos which implies that $\mstarmax$ is likely to be larger as well. Hence, the value of $r_\rmP$ contains valuable information regarding the scatter in the (typically unknown) halo masses of the host galaxies\footnote{We have verified that the non-negligible amount of correlation in the $S_0$ sample is mainly driven by ``extreme value" statistics where, for a fixed underlying SHMF, the act of drawing more satellites ($\nsat$) increases your chances of selecting a higher stellar mass ($\mstarmax$).}. This information is entirely lost when using the stacked SSMF to constrain the model. Another important point to make is that the scatter in the value of $r_\rmP$ for a given sample is appreciable. This indicates that a survey size of 100 host galaxies is not sufficient to beat down sample variance. This also explains why we use a staggering $10,000$ host halo realizations for our modeling, as we seek to suppress sample variance in our model predictions. As
detailed in Appendix~\ref{app:convergence}, such a large number of host halos is indeed required to make reliable model predictions.

\subsection{Statistical Framework}
\label{sec:stats}

Consider a SAGA-like survey. Let $n_{{\rm sat},i}$ be the number of observed satellite galaxies around host $i$ ($i=1,\Nhost$), and let $M_{\ast, i}^{\rm max}$ be the stellar mass of its most massive satellite. Unless specifically mentioned otherwise, we adopt $N_{\rm host}=100$. We now specify how, given this data, we constrain the parameters that specify the SHMR (see Section~\ref{subsec:SHMR}). Let $\bD = (n_{{\rm sat},1}, n_{{\rm sat},2},...,n_{{\rm sat},\Nhost},M_{\ast,1}^{\rm max},M_{\ast,2}^{\rm max},...,M_{\ast,\Nhost}^{\rm max})$ be the data vector, and let $\btheta = (\alpha, \sigma)$ be the vector characterizing our model for the SHMR. We seek to compute the posterior $P(\btheta|\bD)$. From Bayes theorem
\begin{equation}
P(\btheta|\bD) \propto \calL(\bD|\btheta) \, P(\btheta)
\end{equation}
where $P(\btheta)$ is the prior probability distribution for the model parameters, and $\calL(\bD|\btheta)$ is the likelihood of the mock data given the model. Throughout we use uniform, non-informative priors for all of our model parameters, with ranges large enough such that the posterior is not prior limited.

For a given model, characterized by the vector $\btheta$, we use the 10,000 merger trees that were not used in the construction of the mock data to construct satellite populations around $\Nmodel = 10,000$ host halos. Let $\hat{n}_{j}$ be the number of satellite galaxies around host $j$ ($j=1,\Nmodel$), and let $\hat{M}_{\ast,j}^{\rm max}$ be the stellar mass of its most massive satellite. Here $\hat{.}$ is used for clarity to indicate that this quantity refers to the model, rather than the (mock) data. 

Using the summary statistics described in Section~\ref{subsec:data}, we have that
\begin{equation}\label{totlikelihood}
\ln\calL(\bD|\btheta) = 
\ln\calL_N(\bD|\btheta)+ \ln\calL_{M|N}(\bD|\btheta) 
\end{equation}
with $\calL_N(\bD|\btheta)$ the likelihood for the observed number of satellites for each host given the model, and $\calL_{M|N}(\bD|\btheta)$ the likelihood for the masses of their most massive satellites conditioned on the number of satellites found in the same host system. We have that
\begin{equation}\label{lnLN}
\ln\calL_N(\bD|\btheta) = \sum\limits_{i=1}^{\Nhost} \ln P(n_{{\rm sat},i}|\btheta)\,.
\end{equation}
where
\begin{align}\label{Pnobs}
P(n_{{\rm sat},i}|\btheta) = \frac{1}{\Nmodel}
\sum_{j=1}^{\Nmodel} \delta(\hat{n}_{j} - n_{{\rm sat},i})
\end{align}
with $\delta(x)$ the Dirac delta function. Hence, the probability that a given host in the SAGA-like survey has $n_{\rm sat}$ satellites is inferred from the frequency distribution of that number of satellites among the $\Nmodel = 10,000$ host realizations (equivalent to 100 individual SAGA-like surveys).

The second term of equation~(\ref{totlikelihood}) expresses the likelihood for the stellar masses of the most massive satellites given the number of satellites in the same host system. Using the set of most massive satellites from the mock data, $\{M_{\ast,i}^{\rm max}\}$, and the set of most massive stellar masses for the 10,000 host halos computed as a representation of model $\btheta$, $\{\hat{M}_{\ast,j}^{\rm max}\}$, we could in principle compute the likelihood of the former using the Kolmogorov–Smirnov ``two sample" test (KS-test), i.e., $\calL_M(\bD|\btheta) = P_{\rm KS}(\{M_{\ast,i}^{\rm max}\},\{\hat{M}_{\ast,j}^{\rm max}\})$, where $P_{\rm KS}$ is the $p$-value returned by the KS-test. However, as shown in Section~\ref{subsec:data}, the stellar mass of the most massive satellite is correlated with the number of satellites, and the strength of this correlation contains important information. In order to take this into account, we proceed as follows. Let $N_l$ be the number of hosts in the mock data for which $n_{\rm sat}=l$, and let $\{M_{\ast,i}^{\rm max}\}_l$ be the subset of the most massive stellar masses in those hosts. An example of such a subset is highlighted by the rectangle in Panel [F] of Fig.~\ref{fig:summary_stats}, which picks out those host galaxies that have $n_{\rm sat}=6$ satellites. Similarly, let $\{\hat{M}_{\ast,j}^{\rm max}\}_l$ be the subset of the most massive stellar masses in those 10,000 model host halos with $\hat{n}_{j}=l$.  We then perform the KS-test separately for each value of $l$ for which $N_l \geq 4$, which we combine to define the likelihood\footnote{Formally, for data vector of length $N_l$ and model vector of length $\hat{N}_l$, the KS test requires that $\frac{N_l \hat{N}_l}{N_l + \hat{N}_l} > 4$ \citep[][]{Press.etal.92}. However, given that in our framework $\hat{N}_l >> N_l$, we only require $N_l > 4$.}
\begin{equation}
\ln\calL_{M|N}(\bD|\btheta) = \sum\limits_{l; N_l\geq 4} \ln P_{\rm KS}(\{M_{\ast,i}^{\rm max}\}_l,\{\hat{M}_{\ast,j}^{\rm max}\}_l) 
\end{equation}
where the summation is over all $l$ for which $N_{\rm l} \geq 4$.

To summarize, for a given model $\btheta$ we construct satellite populations for $\Nmodel = 10,000$ host halos which we use to approximate $P(\nsat)$, the probability that a host in the SAGA-like survey contains $\nsat$ satellites, and the probability $P(\mstarmax|\nsat)$ that the most massive satellites in a host with $\nsat$ satellites has mass $\mstarmax$. By considering the latter probability function conditioned on the number of satellites, we account for the correlation between $\nsat$ and $\mstarmax$ discussed in Section~\ref{subsec:data}. 

The fact that we only use two scalar quantities to characterize the entire satellite population per host begs the question whether including more information might be valuable to the analysis.  In principle, one can extend the method proposed here by also folding in constraints on the $n^{\rm th}$ most massive satellite galaxies, with $n = 2,..,n_{\rm sat}$, or on the total stellar mass of all satellites. However, as discussed in Appendix~\ref{app:stats}, this is far from trivial due to the fact that all these rank-ordered stellar masses are strongly correlated. This needs to be properly accounted for in the modelling, which quickly becomes prohibitively expensive. Hence, we only base our inference on $\nsat$ and $\mstarmax$. As shown in Appendix~\ref{app:stats}, we find that using the total stellar mass of all satellites, rather than only that of the most massive satellite, yields results that are virtually indistinguishable. This suggests, but doesn't prove, that adding information on the stellar masses of the $n^{\rm th}$ most massive satellites is unlikely to yield constraints that are significantly tighter. This is at least consistent with the notion that the cumulative stellar mass functions of satellite galaxies around individual host galaxies are ``self-similar" in that they can be quantified by two numbers; a normalization ($n_{\rm sat}$) and a characteristic mass ($\mstarmax$). This has a natural explanation in the fact that the underlying subhalo mass function is also self-similar \citep[][]{Jiang.vdBosch.16}.

Finally, to sample the posterior, $P(\btheta|\bD)$, we use the affine invariant ensemble sampler \citep[][]{Goodman.Weare.10} in the software package \textsc{emcee} \citep[][]{Foreman-Mackey.etal.13}. Throughout we set the ``stretch-move" parameter to its default value ($2.0$), which yields acceptance fractions in the range from 25\% to 40\%, depending on the model, and we run the chains until convergence making use of \textsc{emcee}'s build-in autocorrelation tester.

\section{Results}
\label{sec:res}

We now investigate to what level of accuracy and precision our mock surveys can be used to constrain the low mass end of the SHMR. We first pick a model, $\btheta_{\rm true}$, and use the first 100 of the 10,100 merger trees of one of our three samples ($S_0$, $S_{15}$ or $S_{30}$) to construct a mock SAGA-like sample of satellites, assumed to be complete down to a limiting stellar mass of $\mlim$. Next, we use the methodology outlined in Section~\ref{sec:stats} to constrain the posterior $P(\btheta|\bD)$. Importantly, in this section we only consider models with the same degrees of freedom as the model used to construct the mock data. In other words, we work under the hypothesis that the model is a perfect description of reality and that we have perfect knowledge of the distribution of host halo masses. In Sections~\ref{subsec:extra_freedom} and \ref{sec:mass_mixing} we will relax this assumption and investigate how an incorrect, or incomplete model, introduces systematic errors in the inference.

We start our investigation by considering a simple two-parameter SHMR model, characterized by the power-law slope, $\alpha$, and a constant (mass-independent) scatter, $\sigma$. Next, in Section~\ref{subsec:extra_freedom}, we examine how well we can detect deviations from such a simple model in the form of mass dependence in the scatter or curvature (with and without redshift dependence). We also examine how the level of inference depends on the size ($\Nhost$) and depth ($\mlim$) of the survey Section~\ref{subsec:survey_limits}). First, though, we develop important insight by highlighting the various sources of scatter that contribute to the substantial host-to-host variance in a SAGA-like survey.

\subsection{Sources of Scatter}
\label{subsec:scatter}

The host-to-host variance in the number of satellites, $\nsat$, and the mass of the most massive satellite, $\mstarmax$, in a satellite galaxy survey has three distinct sources: halo-to-halo variance in the merger histories of halos of a given mass, variance in the masses of the host halos, and scatter in the SHMR (as characterised by equation~[\ref{sigma_M}]). Here we examine which of these sources dominates the host-to-host\footnote{``Host-to-host" variance is a combination of all three sources of scatter and should not be confused with ``halo-to-halo" variance which refers only to the variance caused by the differing accretion histories of host halos with the {\it same} mass} variance in $\nsat$ and $\mstarmax$. To do so, we proceed as follows. We first use the $S_{0}$ sample to construct 100 mock SAGA-like surveys, each with 100 host halos with a mass of $M_\rmH = 10^{12}\Msun$. Each host halo is populated with satellite galaxies using our fiducial power-law SHMR with $\alpha = 2.0$ but with zero scatter (i.e., $\sigma = 0.0$). For each mock SAGA-like survey we then compute the variance $\sigma_{\nsat}$ in the number of satellites and the variance $\sigma_{\mstarmax}$ in $\log(\mstarmax)$.

\begin{figure}
    \centering
    \includegraphics{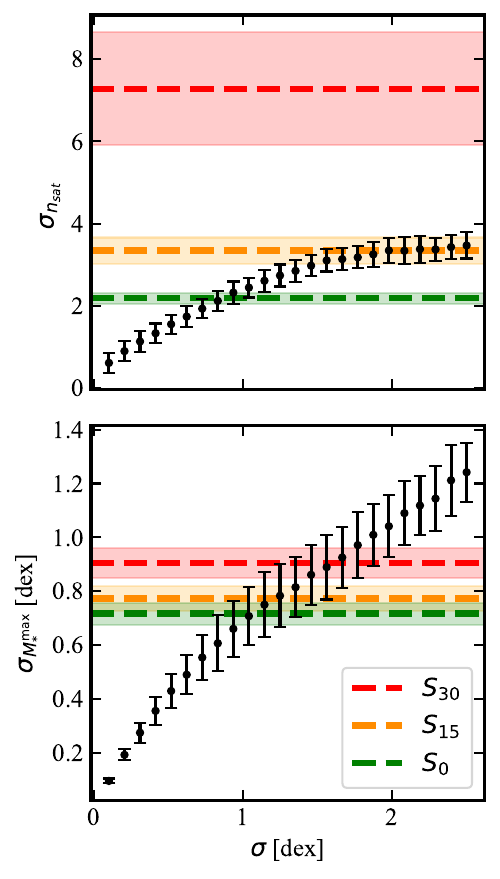}
    \caption{Sources of host-to-host variance in $\nsat$ (top panel) and $\log(\mstarmax)$ (bottom panel) in a SAGA-like survey. The green shaded region (dashed line) indicate the $16-84$ percentile range (median) in a survey with zero scatter in host halo masses and zero scatter in the SHMR ($\sigma = 0$), indicating the scatter due solely to variance in halo MAHs. Horizontal bands in yellow and red show the same but assuming $0.15$dex and $0.3$dex scatter in host halo masses, respectively. Finally, the black dots indicate the average host-to-host variance (with $1\sigma$ errorbars) due solely to scatter in the SHMR as a function of $\sigma$. Unless the latter exceeds $\sim 1$dex, which is unlikely, the host-to-host variance in satellite populations clearly dominated by variance in the MAHs of the host halos.}
    \label{fig:scatter}
\end{figure}

The green horizontal bars in Fig.~\ref{fig:scatter} show the $16-84$ percentile ranges among the 100 mock surveys thus obtained, and are indicative of the scatter arising solely from the halo-to-halo variance in the MAHs for halos of a given mass.  Next we repeat the same exercise for the $S_{15}$ and $S_{30}$ samples, the results of which are shown as the orange and red bars, respectively. These show the combined impact of scatter arising from MAH variance and the sampling of different host halo masses, again in the absence of any scatter in the SHMR. Note how even a modest $0.3$~dex scatter in host halo mass drastically boosts the host-to-host variance in the satellite populations. This underlines that it is crucial that any reliable assessment of the SAGA (or ELVES) data carefully marginalizes over the uncertain distribution of host halo masses. 

Finally, we randomly pick one of the 10,000 merger trees from sample $S_0$ for which we make 100 realizations of satellites galaxies, each time drawing stellar masses from our fiducial power-law SHMR for a specific value of $\sigma$, and computing the scatter in $\nsat$ and $\log(\mstarmax)$. We then repeat this exercise 100 times for 100 different randomly selected merger trees from sample $S_0$. The black symbols with errorbars in Fig.~\ref{fig:scatter} show the resulting $\sigma_{n_{\rm sat}}$ and $\sigma_{\mstarmax}$ as a function of the scatter in the SHMR: $\sigma$. These show the variance in $\nsat$ and $\mstarmax$ due solely to scatter in the SHMR. Note how $\sigma_{\nsat}$ increases monotonically with the increasing amount of scatter in the SHMR but saturates at $\sigma \approx$ 1.5~dex.   

As is evident, the scatter due to halo-to-halo variance in merger histories clearly dominates over that due to the scatter in the SHMR unless the latter is extremely large. Even in the unrealistic case of zero scatter in host halo masses (i.e., the $S_0$ sample), the scatter in the SHMR needs to exceed almost 1dex in order to become the dominant source of scatter responsible for the host-to-host variance. If current hydrodynamical simulations of galaxy formation are an indication, such a large amount of scatter is not to be expected \citep[e.g.,][]{Garrison-Kimmel.etal.17, Fitts.etal.18, Munshi.etal.21, Sales.etal.22}. Hence, we should expect that constraining the scatter in the SHMR is extremely challenging, unless one has a survey large enough to beat down the host-to-host variance.

\subsection{A power-law SHMR}
\label{subsec:powerlaw}

\begin{figure*}
    \centering
    \includegraphics{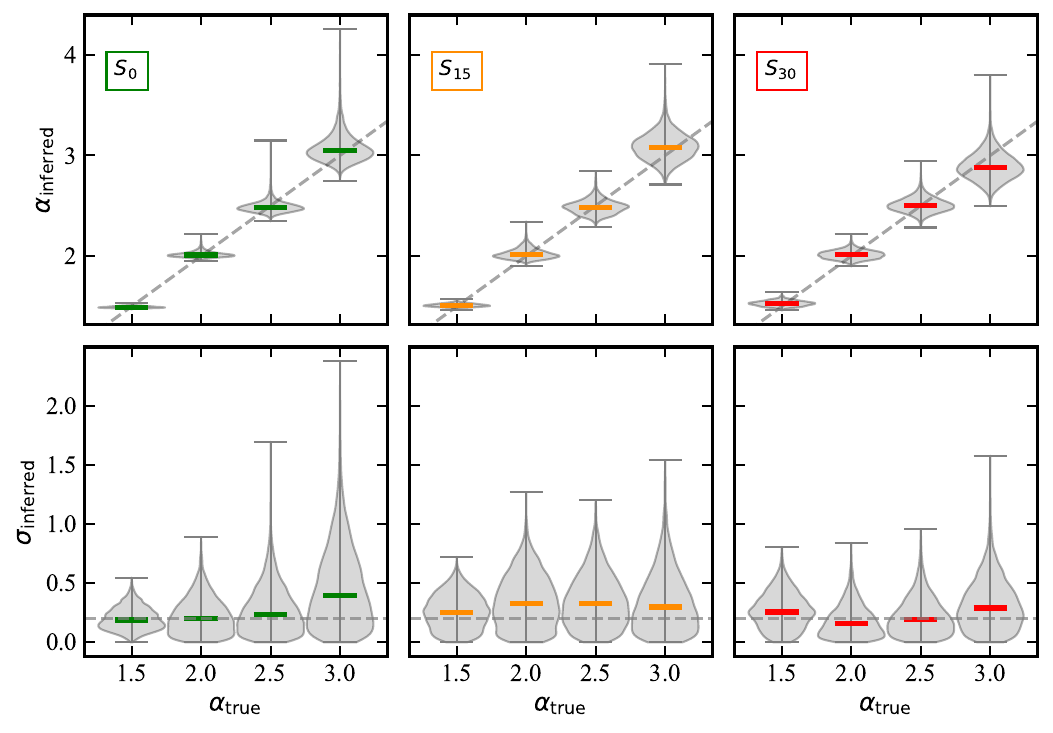}
    \caption{Violin plots depicting the posterior constraints on the power-law slope, $\alpha$, of the SHMR. Each panel shows the results for four different mock samples that only differ in the true input value of $\alpha$ used to construct the mock. Each mock is constructed assuming a mass-independent scatter in the SHMR of $\sigma=0.2$dex. The posterior constraints on $\sigma$, which are also fit for, are shown on the bottom row. Different panels correspond to different samples, as indicated. Finally, the dashed line in each panel indicates the one-to-one correspondence. See Table~\ref{tab:alpha} for the corresponding confidence levels.}
    \label{fig:alpha_permutations}
\end{figure*}

\begin{figure*}
    \centering
    \includegraphics{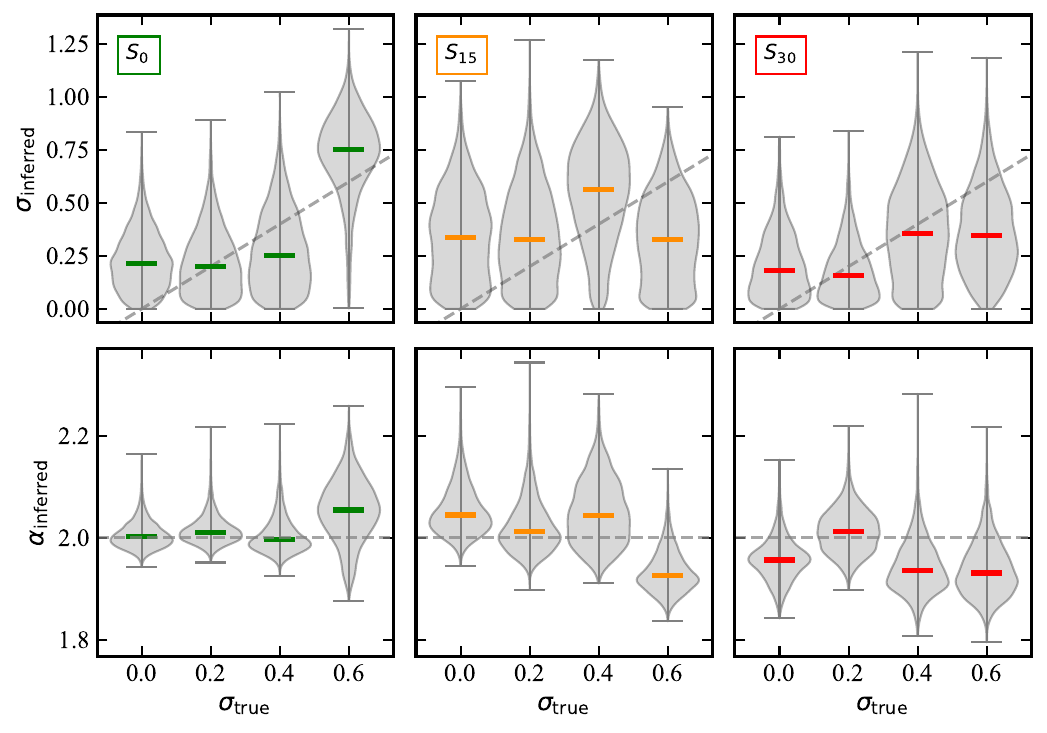}
    \caption{Violin plots depicting the posterior constraints on the scatter in stellar mass, $\sigma$, of the SHMR. Each panel shows the results for four different mock samples that only differ in the true input value of $\sigma$ used to construct the mock. Each mock is constructed assuming a mass-independent scatter in the SHMR of $\alpha=2.0$. The posterior constraints on $\sigma$, which are also fit for, are shown on the bottom row. Different panels correspond to different samples, as indicated. Finally, the dashed lines in each panel mark the one-to-one correspondence. See Table~\ref{tab:sigma} for the corresponding confidence levels.}
    \label{fig:sigma_permutations}
\end{figure*}

We start our investigation by adopting the simple power-law model for the SHMR, characterized by the power-law slope, $\alpha$, and a mass independent scatter, $\sigma$. This is the model that has been used almost exclusively in previous studies, and therefore is a logical starting point We will consider more generic models in Section~\ref{subsec:extra_freedom} below. 

For each of our three samples ($S_0$, $S_{15}$ and $S_{30}$), we construct 7 different models: the fiducial model with $(\alpha,\sigma)=(2.0,0.2)$, three versions with $\sigma=0.2$ but different values of $\alpha$ ($1.5$, $2.5$, and $3.0$), and three versions with $\alpha=2.0$ but different values for the scatter ($\sigma = 0.0$, $0.4$, and $0.6$).  Each of these mock data samples is subsequently analyzed using the method outlined above. Typically, we first run a coarse grid over the prior parameter space, and then start the MCMC at the grid-position for which the likelihood is largest. Final posteriors are sampled from the chains using a standardized thinning algorithm. Results for all these models, and all other models discussed in this paper, are summarized in table-form in Appendix~\ref{app:results}.

The upper panels of Fig.~\ref{fig:alpha_permutations} show violin-plots depicting the posterior constraints on $\alpha$ for the four models with $\sigma = 0.2$. From left to right, the panels correspond to results for the $S_0$ sample (no scatter in host halo mass), the $S_{15}$ sample (0.15~dex scatter in halo mass), and the $S_{30}$ sample (0.3~dex scatter in host halo mass). Thick, colored horizontal bars depict the median values of the posterior distributions, while the grey errorbars indicate their full extent.  The dashed, diagonal line in each upper panel marks the one-to-one correspondence and is shown for ease of comparison. The following trends are evident: (i) overall, the inferred value of the power-law slope is in good agreement with the assumed input value, (ii) the inferred value of $\alpha$ is more accurate and precise if its true value is smaller (i.e., the SHMR has a shallower slope), and (iii) the inference is not dramatically impacted by scatter in host halo mass, at least if its distribution is known, as is assumed here.

The fact that the inference improves with decreasing $\alpha$ is simply a consequence of the fact that a shallower power-law slope implies more satellites. For example, the average number of satellites per MW-like host in the sample $S_{15}$ decreases from $\langle \nsat\rangle \approx 32$ for $\alpha=1.5$, to $\langle \nsat\rangle \lesssim 1$ for $\alpha=3.0$ (these averages are obtained from 100 independent mock samples).

The lower panels of Fig.~\ref{fig:alpha_permutations} show violin-plots depicting the corresponding posterior constraints on the scatter, $\sigma$. The dashed, horizontal lines indicate the true input value, $\sigma=0.2$. Although the medians of the posterior distributions do, in most cases, agree with this true input value, overall the constraints on $\sigma$ are extremely poor. Best case scenario, one only infers that $\sigma \lta 0.5$dex at 95\% confidence level (CL). Given that the scatter in the SHMR at the massive end ($M_\rmH \gta 10^{11} \Msun$) is constrained to be 0.15-0.2dex (see Section~\ref{sec:intro}), such constraints are not very meaningful.

Fig.~\ref{fig:sigma_permutations} is similar to Fig.~\ref{fig:alpha_permutations} but this time we show the results for the four mock data sets with $\alpha=2.0$, but with different values of $\sigma$. Once again, different columns correspond to different samples (i.e., different amounts mass mixing), as indicated. We can again clearly see that the posterior constraints on the inferred scatter are extremely poor and show no correlation with the true input value or with the amount of scatter in host halo mass. Hence, as already anticipated in Section~\ref{subsec:scatter}, a SAGA-like survey will not be able to put meaningful constraints on the scatter in the SHMR, even if the SHMR is characterized as a simple power-law with constant scatter, all host halos have identical mass, and incompleteness effects, errors in stellar mass, and interloper issues are absent.
\subsection{Curvature, Mass and Redshift Dependence}
\label{subsec:extra_freedom}

\begin{figure*}
    \centering
    \includegraphics{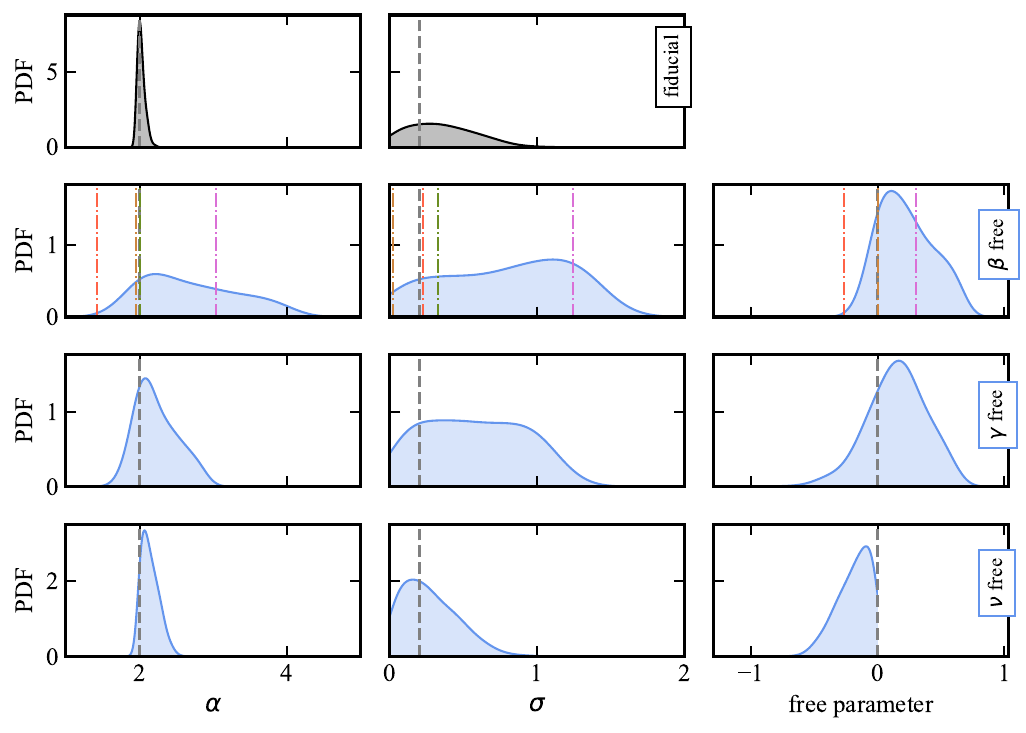}
    \caption{Posterior constraints from four different analyses of the same mock survey constructed using the $S_{15}$ sample and our fiducial SHMR. The left and middle column of panels show the constraints on $\alpha$ and $\sigma$, respectively, while the right column shows the constraints on a third parameter that is kept free. The top row corresponds to the analysis in which only $\alpha$ and $\sigma$ are allowed to vary. The other rows, from top to bottom, show the results from analyses in which $\beta$ is left free, $\gamma$ is left free, and $\nu$ is left free. The dashed grey vertical lines show the true input value used to construct the mock data. The four vertical colored lines in the second row from the top (``$\beta$ free") indicate an illustrative selection of models shown in Fig.~\ref{fig:model_illustration}. See text for a detailed discussion and Table~\ref{tab:free} for the corresponding confidence levels.} 
    \label{fig:extra_freedom}
\end{figure*}

\begin{figure*}
    \centering
    \includegraphics{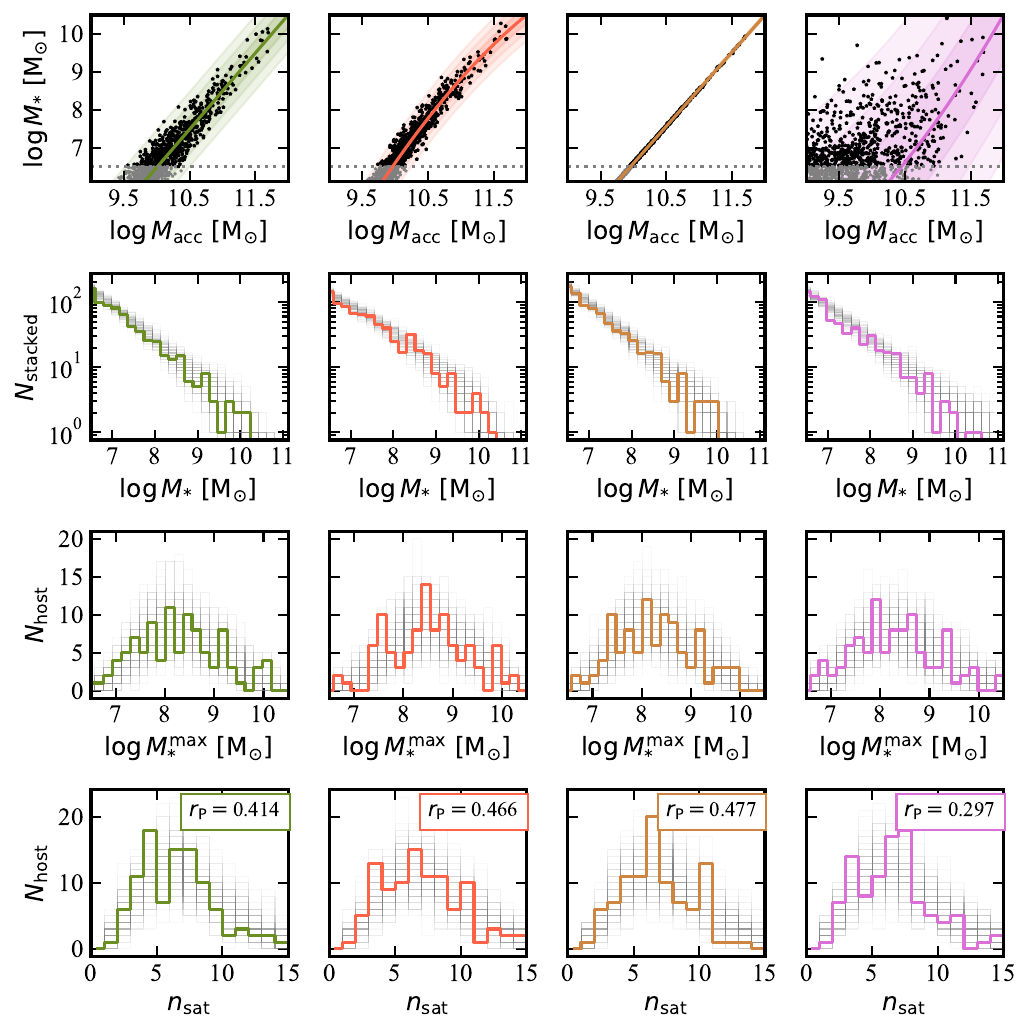}
    \caption{An illustrative selection of four wildly different SHMRs, taken from the``$\beta$ free" inferred posterior distribution, and their associated summary statistics. Each column represents a different parametrization of the SHMR and its color matches that of the corresponding, vertical dashed line in the second row of Fig.~\ref{fig:extra_freedom}. The top rows show the actual SHMR (solid lines), the 1, 2 and $3\sigma$ scatter intervals (shaded contours), and a single mock survey realization (black asterisks) constructed using the $S_{15}$ sample. The second row shows the corresponding “stacked” SSMFs as thick colored line (cf. panel [C] in Fig.~\ref{fig:summary_stats}). The third and bottom rows shows histograms of the $\mstarmax$ and $\nsat$ respectively (cf. panels [D] and [E] in Fig.~\ref{fig:summary_stats}, respectively) along with the corresponding Pearson’s rank-order correlation coefficient. In the background of the bottom three rows we show the ``SAGA-to-SAGA" variance in the summary statistics as thin grey lines. Despite the wildly different SHMRs, these four mock SAGA-like surveys are virtually indistinguishable in terms of their summary statistics.}
    \label{fig:model_illustration}
\end{figure*}

Thus far we have assumed that the SHMR is characterized by a simple power-law with mass independent scatter. However, there is no a-priori reason why the intricate process of galaxy formation cannot result in a more complicated functional form for $\mstar(M_\rmH)$. We therefore adopt the following, more flexible version of equation~(\ref{SHMR_PL}):
\begin{align}\label{SHMR}
 \log[\mstar] & = \log[\mstaranchor] + \alpha \log[M_{12}] + \beta \log^2[M_{12}] \nonumber \\
 &+ \calG[\sigma(\macc)].
\end{align}
Here $\beta$ describes a quadratic deviation from the simply power-law, which we allow to have a redshift dependence given by 
\begin{equation}\label{beta_z}
\beta(\zacc) = \beta + \gamma \log[1 + \zacc]\,.
\end{equation}
This is motivated by the work of \citet[][]{Lu.etal.14, Lu.etal.15} who have shown that a model with such a redshift dependence can explain the low-mass end of the galaxy stellar mass function, especially at higher redshifts.  We also allow for a halo mass dependence in the scatter in the SHMR, according to 
\begin{equation}\label{sigma_M}
\sigma(\macc) = \sigma + \nu \log[M_{12}]\,,
\end{equation}
which adds another free parameters, $\nu$, which characterizes how the scatter scales with halo mass. This is motivated by the work of \cite[][and references therein]{Munshi.etal.21} who argue that the scatter in the SHMR likely increases towards lower halo masses. With this additional freedom, the SHMR model is characterized by a total of five free parameters; $\btheta = (\alpha, \beta, \gamma, \sigma, \nu)$. For simplicity, however, we do not allow all five parameters to vary at once. Instead, we only compare how each parameter in turn alters the inference with respect to the simple power-law model.

We construct a mock SAGA-like survey from our $S_{15}$ sample of merger trees using the fiducial SHMR model with $\thetafid$. We then analyze this mock data using four different models: a fiducial model, in which only $\alpha$ and $\sigma$ are allowed to vary while the other three parameters are kept fixed at zero, and three modifications thereof; a model in which $\beta$ is allowed to vary, a model in which $\gamma$ is allowed to vary, and a model in which $\nu$ is allowed to vary. In this way we compare our fiducial model with three different models that each explore one extra degree of freedom. Note that all models use the 10,000 merger trees from the $S_{15}$ sample that were not used in the construction of the mock data. The resulting posterior constraints are shown in Fig.~\ref{fig:extra_freedom}. Different rows correspond to the four different models used for the analysis. The left and middle column show the posterior constraints on $\alpha$ and $\sigma$, respectively, while the right column shows the constraints on the additional 
parameter that is kept free. 

Comparing the first two rows, it is evident that allowing for curvature, $\beta$, in the SHMR largely erases the constraining power on the power-law slope $\alpha$. The constraints change from $\alpha=2.01_{-0.05}^{+0.11}$ (95\% CL) when $\beta$ is kept fixed at zero to $\alpha=2.59_{-0.74}^{+1.27}$ when $\beta$ is kept free. Note that the constraints on $\sigma$ also weakens significantly, while $\beta$ itself is only poorly constrained. In fact, although not explicitly shown in Fig.~\ref{fig:extra_freedom} we find that $\alpha$ and $\beta$ are strongly degenerate, with larger values of $\alpha$ implying larger positive curvature. 

The third row in Fig.~\ref{fig:extra_freedom} shows that allowing for redshift evolution in the curvature parameter (but restricting the curvature to be zero at $z=0$) once again significantly weakens the constraints on $\alpha$. And since the data is unable to constrain the curvature itself, it shouldn't come as a surprise that it is also unable to meaningfully constrain its redshift evolution. Indeed, we find that $\gamma$ can only be constrained to $0.16_{-0.405}^{+0.365}$ at the 95\% CL.

Finally, the bottom row in Fig.~\ref{fig:extra_freedom} shows that the mock survey is unable to put any meaningful constraints on the halo mass dependence of the scatter in the SHMR. Note that we only allow for negative $\nu$ (i.e., scatter increases with decreasing halo mass) While the mock data has $\nu=0$ (no mass dependence), $\nu$ is constrained to $-0.16_{-0.276}^{+0.148}$ at the 95\% CL. This indicates once more that a SAGA-like survey has little to no constraining power regarding the scatter in the SHMR.

To highlight the difficulty constraining the SHMR using satellite galaxy populations, Fig.~\ref{fig:model_illustration} compares the SHMRs and various summary statistics for four wildly different models (each based on and analyzed with the $S_{15}$ sample). Each of these models are taken from the ``$\beta$ free'' posteriors shown as dotted, colored lines in the second row of Fig.~\ref{fig:extra_freedom}. As is evident from the top panels, these SHMRs differ dramatically from each other. Yet, the lower three panels clearly show that the summary statistics discussed in this paper (i.e., the total stellar mass function and the distributions of $n_{\rm sat}$ and $M_{\ast}^{\rm max}$) are virtually indistinguishable. Note also that the Pearson rank-order correlation coefficients for the different models (indicated in the panels in the top row) are all consistent with the distribution of $r_\rmP$ values expected for the fiducial $S_{15}$-based model shown (in orange) in Fig.~\ref{fig:correlation}. This clearly underscores the challenges one has to overcome when using satellite surveys to constrain the SHMR.

\subsection{Dependence on Survey Size and Depth}
\label{subsec:survey_limits}

Thus far we have assumed that the survey is complete down to a limiting stellar mass of $\mlim = 10^{6.5}\Msun$. However, the real SAGA survey is only $<34\%$ complete down to this stellar mass limit \citep[][]{Mao.etal.24}. The survey is $94\%$ complete, though, down to a stellar mass of $10^{7.5}$ \citep[referred to as the Gold sample in][]{Mao.etal.24}. For comparison, the ELVES survey claims to reach down to $\sim 10^{5.5}\Msun$ albeit with $\lesssim$ 50\% completeness \citep[][]{Danieli.etal.23}. Here we test how sensitive the SHMR constraints are to the limiting stellar mass of the survey, $\mlim$. Throughout we continue to assume $100\%$ completeness down to $\mstar = \mlim$. 

\begin{figure*}
    \centering
    \includegraphics{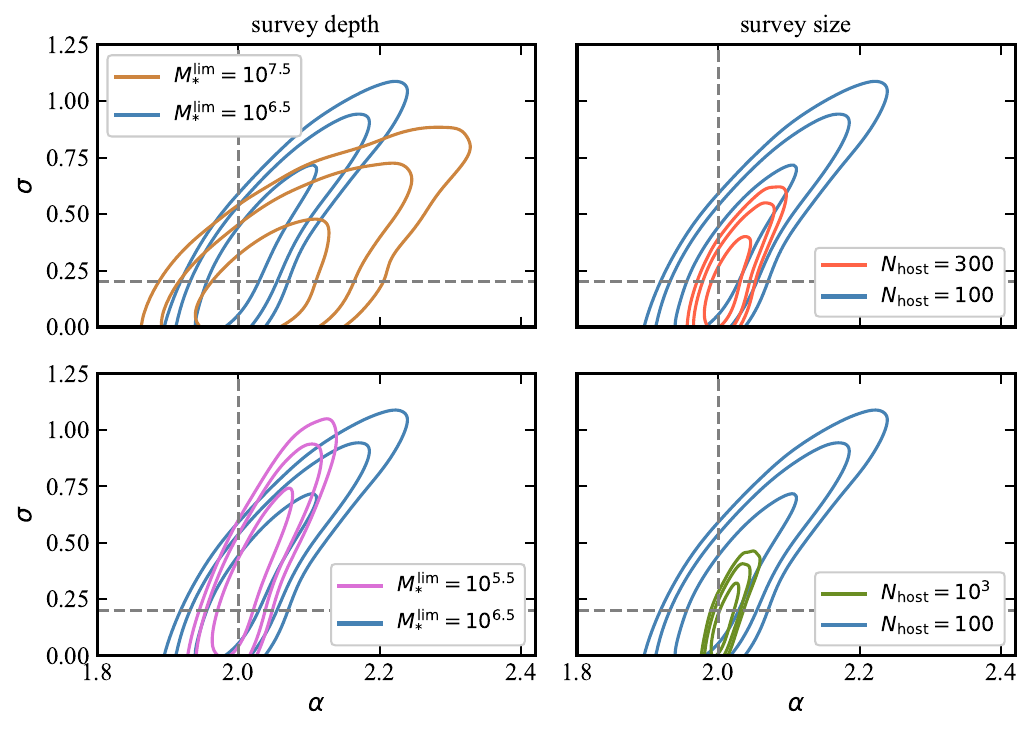}
    \caption{The inferred posterior distributions for $\alpha$ and $\sigma$ for different survey depths (left-panel) and sample sizes (right-panel) in a satellite galaxy survey. The grey dashed lines indicate the true ``input" values ($\thetafid$) used to construct the mock data. See text for discussion and Table~\ref{tab:future} for the corresponding CL.}
    \label{fig:survey_limits}
\end{figure*}

We construct three mock SAGA-like surveys (each with $\Nhost=100$) from our $S_{15}$ sample adopting the fiducial SHMR with a power-law slope $\alpha=2.0$ and a mass-independent scatter $\sigma=0.2$. These mock surveys only differ in their limiting stellar mass, having $\log[\mlim/\Msun] = 5.5$ (ELVES-like), $6.5$ (our fiducial value) and $7.5$ (the SAGA Gold sample). Next, we analyze these mocks using the 10,000 merger trees from the $S_{15}$ sample that were not used in constructing the mock data, and adopting the fiducial SHMR model (i.e., only $\alpha$ and $\sigma$ are allowed to vary, while $\beta=\gamma=\nu=0$). Results are shown in the left column of panels in Fig.~\ref{fig:survey_limits}. As is apparent, going down to lower stellar mass helps to tighten the constraints on $\alpha$ somewhat, but the constraints on $\sigma$ are largely independent of $\mlim$. 

Rather than aiming for a deeper survey (i.e., complete down to a smaller stellar mass), one could also envision increasing the number of hosts included. In order to assess the merit of a larger sample, we construct two new mock surveys with $\Nhost=300$ and $\Nhost=1000$ (i.e., 3 and 10 times larger than the current SAGA survey, respectively). We again assume 100 percent completeness down to $\mlim = 10^{6.5}$. The mock data is constructed from the $S_{15}$ sample using the fiducial SHMR, and analyzed in the same way. Results are shown in the right column of panels in Fig.~\ref{fig:survey_limits}. As is apparent, increasing the sample size by an order of magnitude can drastically reduce the degeneracy between $\alpha$ and $\sigma$, and has the potential to constrain the scatter to a meaningful range. Hence, we conclude that increasing the sample size by an order of magnitude has more merit than pushing the survey to a limiting stellar mass that is an order of magnitude lower than the current limit.

\section{The Effect of Host Halo Mass Mixing}
\label{sec:mass_mixing}

Up until this point we have always analyzed the mock data using the same sample of merger trees as that used to make the mock data. Hence, we have assumed that we have perfect knowledge, in a statistical sense, of the distribution of host halo masses. In reality though, the distribution of host halo masses is poorly known. In this section we investigate the consequence of over- or under-estimating the variance in host halo masses in a survey. To that extent, we construct three mock SAGA-like surveys using merger trees drawn randomly from samples $S_0$, $S_{15}$ and $S_{30}$, respectively, and using the fiducial SHMR model ($\alpha=2.0$ and $\sigma=0.2$). Next we analyze each of these three mock data sets using our fiducial model (i.e., only $\alpha$ and $\sigma$ are allowed to vary) with each of the three different sets of 10,000 merger trees, taken either from sample $S_{0}$, $S_{15}$ or $S_{30}$. 

\begin{figure*}
    \centering
    \includegraphics{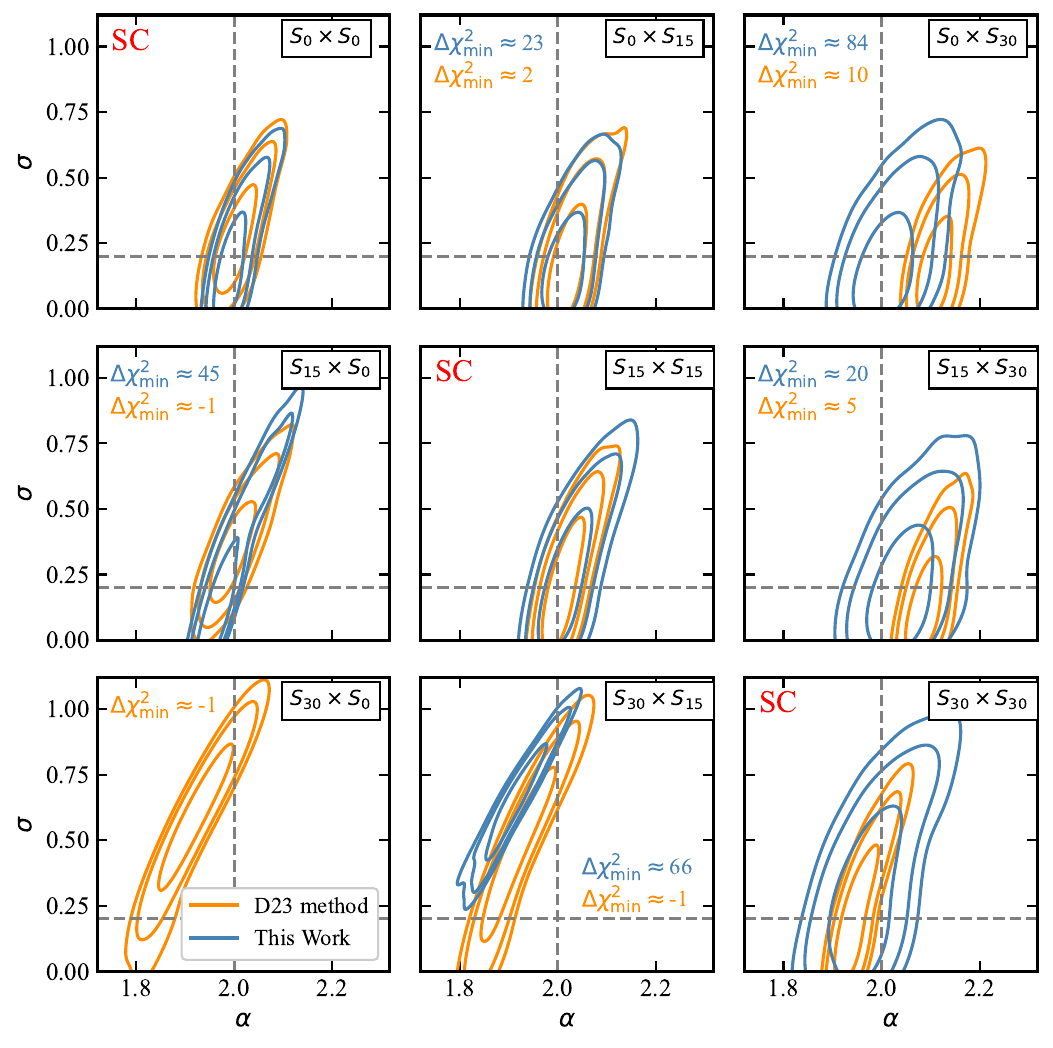}
    \caption{The posterior distributions used to investigate how host halo mass mixing effects the inference. The rectangular box in the top-right corner of each panel indicates the sample used to create the mock data, 'D', and the sample used to model the data, `M' in the format D$\times$M. Hence, different rows correspond to different samples used for the construction of the mock data, while different columns reflect different samples used in their analysis. The panels that run diagonally from the top left to bottom are ``self-consistent'' (SC) in that the same sample of host halos was used in the analysis as was used to construct the mock data. Blue contours indicate the posterior distributions (68, 95 and 99\% CL) obtained using our statistical framework. For comparison, we over-plot in orange the posterior distributions obtained by fitting the combined SSMF using the methodology outlined in D23 (using the likelihood $\calL_{\rm SSMF}(\bD| \theta)$ given by eq.~[\ref{calLelves}]). In the off-diagonal panels, we indicate values of $\Delta \chi^2_{\mathrm{min}}$ for each method (colored accordingly). Here $\Delta \chi^2_{\mathrm{min}} = \chi^2_{\mathrm{min}} - \chi^2_{\mathrm{min, SC}}$ which is the difference in $\chi^2$ between the ``best-fit" model and its corresponding ``best-fit" self-consistent model. The grey dashed lines indicate the true input values of the fiducial SHMR model used to construct the mocks.  The lack of blue contours in the lower-left panel indicates that no model was able to fit the data within the priors. See text for detailed discussion.}
    \label{fig:matrix}
\end{figure*}

The posterior constraints on $\alpha$ and $\sigma$ from these $3 \times 3$ analyses are shown in  Fig.~\ref{fig:matrix}. Here, different rows correspond to different samples used for the construction of the mock data, and different columns correspond to different samples used in the analysis. Hence, panels along the diagonal running from the upper-left to bottom-right are ``self-consistent" (SC) in that the PDF of host halo masses used for the construction of the mock is identical to that used in the analysis (this is identical to what we have done in the previous section). The off-diagonal panels shows the results one obtains if analyzing the data with an incorrect prior on the distribution of host halo masses. Panels below the diagonal correspond to underestimating the amount of mass mixing that is present in the data. Similarly, panels above the diagonal correspond to overestimating the variance in host halo masses.

To elucidate the advantage of our statistical framework over that of D23 (see Section~\ref{subsec:data}), we over-plot two sets of inferred posterior distributions in Fig~\ref{fig:matrix}. Here blue contours indicate the 68, 95 and 99 confidence levels inferred using our methodology, while the orange contours indicate the same confidence levels obtained using an MCMC analysis based on the likelihood, $\calL_{\rm SSMF}$, given by equation~(\ref{calLelves}). Focusing on the blue contours first, one notices that in all but two panels, the inferred posterior distributions are consistent with the input values (grey dashed lines) within the 68\% CL. This indicates that our method typically yields unbiased constraints. The two panels where this is not the case correspond to analyses that underestimate the variance in host halo masses (panels below the diagonal). Since these mock data sets typically contain several host systems with uncharacteristically large abundances of satellites, and our method is based on fitting the entire $\nsat$ distribution, our parameter inference becomes problematic. In fact, for the most extreme case ($S_{30}\times S_{0}$), our analyses completely fails in that no model can fit the data within our priors. Importantly, this failure is to be seen as an asset of our methodology, as it signals a clear shortcoming of one or more of the priors (in this case, the prior on the distribution of host halo masses). 

This is also evident from the corresponding $\Delta \chi^2_{\mathrm{min}}$ values shown in the off-diagonal panels (colored according to which statistical framework was used). Here $\Delta \chi^2_{\mathrm{min}}$ is the difference in $\chi^2$ between an off-diagonal panel's ``best-fit" model and the ``best-fit" model from the SC panel. Note how the blue $\Delta \chi^2_{\mathrm{min}}$ are clearly offset to larger values ($\Delta \chi^2_{\mathrm{min}} \gtrsim 20$) in the off-diagonal panels. This indicates that these off-diagonal inferred posterior distributions can easily be discriminated against. Hence, a researcher will be able to simultaneously constrain the SHMR {\it and} the variance in host halo masses.

The situation is markedly different when using the methodology of D23. As is evident from the orange contours in Fig.~\ref{fig:matrix}, the posterior distributions are systematically offset from the true input values at more than 99\% CL in 5 out of the 9 cases. Yet, the orange $\Delta \chi^2_{\mathrm{min}}$ values are all almost indistinguishable with $\Delta \chi^2_{\mathrm{min}} \lesssim 10$. This is comparable to the range of $\Delta \chi^2_{\mathrm{min}}$ values for the posterior distribution in the self-consistent case. This indicates that just fitting the combined SSMF is unable to diagnose an incorrect prior on host halo masses, which in turn can lead to large, systematic errors in the inference. This is a consequence of the fact that the summary statistic used carries no information about the host-to-host variance in satellite abundance. 

\section{Discussion}
\label{sec:discussion}

In forward modeling the satellite populations of a SAGA-like survey we have made numerous simplified assumptions. Here we briefly discuss some aspects of the modeling that need to be improved upon when using the actual SAGA data to constrain the SHMR. This includes a treatment of observational uncertainties, and a marginalization over the evolution of satellite galaxies post-infall. We close by comparing our findings with similar studies in the literature.

\subsection{Observational Uncertainties}
\label{subsec:uncertainties}

As pointed out in Section~\ref{subsec:mock}, our modeling of SAGA-like surveys ignores effectively all observational uncertainties (and potential biases). In particular, throughout we have assumed the survey data to be 100\% complete down to an arbitrary $\mlim$. In reality, the SAGA survey catalogue is estimated to be $94\%$ ($34\%$) complete down to a stellar mass of $10^{7.5} \Msun$ ($10^{6.75} \Msun$). Given how detrimental sampling fewer satellites per host can be to the inference (see the ``survey depth" test in Section~\ref{subsec:survey_limits}), properly accounting for this mass-dependent incompleteness will be crucial for achieving unbiased constraints on the SHMR.

Another complication ignored here is interlopers (i.e., fore- or background galaxies that are erroneously selected as satellite galaxies in a survey. \citet{Mao.etal.24} estimate the interloper fraction in the SAGA survey to be $\sim$ 30\%. In order to take this into account, one needs to not only model the individual host halos, as done here, but also their immediate environments and the detailed satellite selection criteria. The method used here generates merger histories for one isolated host halo at a time, and is thus ill-suited to model interloper contamination. Cosmological $N$-body simulations are better suited for modeling environments \citep{Christensen.etal.24}, but are much more CPU expensive, suffer from issues related to artificial subhalo disruption, and make it challenging to properly address halo-to-halo variance. 

It is also important to note that we assume zero uncertainty on the ``observed" stellar masses. This is at odds with the real SAGA data, in which the inferred stellar masses have uncertainties of the order of $\sim 0.2$dex. As long as the error distribution is known sufficiently accurately \citep[which is not trivial, see for example][]{Bell.etal.03, Simon.19, Bonfini.etal.21}, it is straightforward to fold in these uncertainties in the forward modeling. To good approximation, observational errors in stellar mass will masquerade as additional scatter in the SHMR. In other words, the constraints on $\sigma$ discussed here should be interpreted as constraints on the quadratic sum of the intrinsic scatter in the SHMR plus the typical error in the observationally inferred stellar masses.

\subsection{Stellar Mass Evolution}
\label{subsec:mstar_evo}

As mentioned in Section~\ref{sec:method}, the inference framework developed here is based on the assumption that the evolution of satellite galaxies is `frozen' the moment they enter the virial radius of a larger host halo. This assumption, which is the essence of subhalo abundance matching \citep[][]{Vale.Ostriker.06, Conroy.etal.06, Guo.etal.10}, is motivated by the idea that satellite galaxies quench their star formation shortly after having been accreted due to the combined effects of ram-pressure stripping \citep[][]{Gunn.Gott.72} and `strangulation' \citep[also known as `starvation'][]{Larson.etal.80, Balogh.etal.00}.

While this assumption facilitates linking what is readily observable (present-day stellar mass) to what is fairly straightforward to model (halo mass at accretion), the notion that a galaxy abruptly stops forming stars once it has been accreted is clearly oversimplified. First of all, it is at odds with the fact that a significant fraction of satellite galaxies are (still) star forming \citep[e.g.,][]{Weinmann.etal.06, Geha.etal.24}. Indeed, numerous studies have shown that there has to be a significant delay time between accretion and quenching, of the order of 1-2 Gyr \citep[e.g.,][]{Wetzel.etal.13, Maier.etal.19, Akins.etal.21}. One possible interpretation is that this delay reflects the time it takes the satellite galaxy to reach its first pericentric passage, where ram-pressure-stripping is most effective at removing gas and quenching the system \citep{Tollet.etal.17}. 

While continued star formation can cause the stellar mass of the satellite to increase post-infall, tidal stripping and tidal heating can have the opposite effect. The importance of these processes is evident from the fact that host galaxies have stellar halos and stellar streams which are the remnants of stripped satellites \citep[][]{Bullock.Johnston.05, Merritt.etal.16, Font.etal.20}. In addition, satellite galaxies have higher metallicities than centrals of the same stellar mass, which is further evidence that (some) satellites may have experienced significant stellar mass loss \citep[][]{Pasquali.etal.10, Gallazzi.etal.21, Oyarzun.etal.23}.

Hence, there are good reasons to assume that the present day stellar mass of an observed satellite galaxy differs substantially from what it was at infall. This needs to be accounted for, or marginalized over, when using the demographics of satellite galaxies to constrain the SHMR. 

\subsection{Subhalo Occupation Fractions and Reionization}
\label{subsec:reionization}

Throughout this work we have made the assumption that each and every subhalo in our merger trees hosts a (satellite) galaxy. However, as eluded to in Section~\ref{sec:intro}, reionization can inhibit galaxy formation in low mass halos, resulting in halo occupation fractions that drop below unity below some mass scale \citep[e.g.,][]{Sawala.etal.16, Fitts.etal.18,  Wheeler.etal.19, Kravtsov.etal.22}. 

The merger trees used in this study all have a mass resolution of $M_\rmH = 10^9 \Msun$. For all reasonable parametrizations of the SHMR, the mock survey's stellar mass limit (modelled here as $10^{6.5}\Msun$) corresponds to a halo mass well above $10^9 \Msun$ (cf., top-left panel of Fig.~\ref{fig:summary_stats}). Although the detailed mass dependence of halo occupation fractions are still relatively uncertain, few if any of the reionization models predict occupation fractions that drop significantly below unity for halos with mass $M_\rmH \geq 10^9 \Msun$. Hence, our results are unlikely to be impacted by the effects reionization. However, whenever using data that goes deeper, it will be important to account for reionization \citep[see for example][]{Jethwa.etal.18, Nadler.etal.24}.

\subsection{Comparison with previous studies}
\label{sec:comp}

One of the main conclusions of this work is that it is extremely difficult to put any meaningful constraints on the scatter in the SHMR using data on the abundances of satellite galaxies. This conclusion appears to be at odds with two recent studies. 

\subsubsection{Nadler et al. (2024)}
\label{sec:N24}

Motivated by upcoming observational facilities that will increase the sample size of MW satellites, \citet[][hereafter N24]{Nadler.etal.24} explores how a {\it complete} sample of MW satellites down to an absolute $V$-band magnitude of $M_V=0$ (a limiting stellar mass of roughly $100\Msun$) can be used to constrain galaxy formation and DM properties. Similar to this work, their analysis revolves around parameter recovery tests. Using the satellite luminosity function as observational constraints, and a suite of 45 cosmological ``zoom-in" dark-matter only simulations of MW-like host halos that span a narrow range ($\sim 0.02$dex) in host halo mass, N24 infer that they can only place an upper limit on the scatter in the SHMR\footnote{In reality, N24 model the scatter in satellite luminosity at given halo mass instead of the scatter in stellar mass. For the purpose of this discussion we assume that the two are directly comparable.} of $\sigma \lta 1.8$dex (95\% CL). However, when folding in data on a potential second host system, with a significantly different satellite abundance, their constraint on the scatter improves drastically to $\sigma = 0.2^{+0.12}_{-0.12}$ (68\% CL), for an input value of $\sigma=0.2$. This appears to be at odds with our conclusion that a SAGA-like survey, which consists of 100 MW-like host systems, can barely constrain the scatter to better than 1.0dex. 

A detailed comparison with the N24 results, however, is difficult and beyond the scope of this work. First of all, the N24 analysis is based on a likelihood analysis that takes as input the numbers of satellites in 14 luminosity bins, whereas our analysis only relies on $\nsat$ and $\mstarmax$. Furthermore, going down to the faintest satellites, as in N24, requires a different set of merger trees with higher mass resolution to resolve subhaloes down to below the reionization scale, where the halo occupation fraction drops below unity.  That being said, if we assume perfect knowledge of the host halo masses (i.e., using the $S_0$ sample), we also obtain tight constraints on $\sigma$ for SHMR slopes that are sufficiently shallow. In particular, for $\alpha=1.5$ we find $\sigma = 0.2^{+0.18}_{-0.12}$ (cf., Table~\ref{tab:alpha}), which is actually very similar to N24. Given the large differences between our strategy and that of N24, this agreement has to be taken with a grain of salt. On the one hand, N24 uses data that extends to much lower luminosities which might improve their constraining power compared to ours. On the other hand, the fact that their analysis only uses $\hat{N}_{\rm host} = 10$ merger trees per model evaluation, compared to our $\hat{N}_{\rm host} = 10^4$, might lead to underestimated uncertainties, as discussed in Appendix~\ref{app:convergence} and Section~\ref{sec:D23} below.

\subsubsection{Danieli et al. (2023)}
\label{sec:D23}

In the recent study by Danieli et al. (D23), the authors use data from the ELVES survey in order to constrain the low mass end of the SHMR. As described in Section \ref{subsec:data}, they combine the satellite populations from all 27 ELVES hosts into a single ``Local Volume Satellite Stellar Mass Function" (LV-SSMF). For their modeling, they use a library of 280 different merger trees, generated using \SatGen, that range in mass from $M_\rmH = 10^{10.5}\Msun$ to $10^{13.3}\Msun$ in steps of 0.01dex. These are then used to construct a ``Local Volume Subhalo Mass Function" (LV-SHMF) as follows. For each host galaxy, the stellar mass of the host, $M_\ast$, is calculated from its $K$-band magnitude assuming a fixed stellar mass-to-light ratio. Next they draw 50 realizations of the host's halo mass, by drawing masses from a log-normal distribution centered on the $z=0$ SHMR, $\langle M_\rmH|M_\ast \rangle$, of \cite{RodriguezPuebla.etal.17}, assuming a constant scatter in halo mass at fixed stellar mass of 0.15~dex. For each of these 50 halo masses they then use the subhalo catalog predicted by the \SatGen-generated merger tree for the one halo (out of 280) that is closest in host halo mass. These $50 \times 27 = 1350$ subhalo catalogs are then combined into a final LV-SHMF.

The final component of their modeling is similar to that presented here. Assuming that the SHMR can be modelled as a power-law relation with slope $\alpha$ and a fixed amount of scatter, $\sigma$, they then assign stellar masses to each subhalo based on its mass at infall. After applying observational selection functions and observational incompleteness corrections applicable to the ELVES survey, the resulting LV-SSMF is then compared to the real data, and a MCMC method is used to constrain the posteriors of $\alpha$ and $\sigma$.

They find $\alpha = 2.10^{+0.01}_{-0.01}$ and $\sigma = 0.06^{+0.07}_{-0.05}$ at the 95\% CL. In light of the results presented here, these constraints are suprisingly tight. For comparison, if we construct a mock SAGA-like survey using a simple power-law SHMR with $\alpha=2.0$ and constant scatter $\sigma=0.2$ using the $S_{15}$ sample of merger histories and analyse it self-consistently (i.e., assuming a power-law SHMR with fixed scatter, and using the $S_{15}$ sample), we obtain as posterior constraints $\alpha = 2.01_{-0.110}^{+0.055}$ and $\sigma =0.33_{-0.405}^{+0.295}$ at the same 95\% CL (see Table~\ref{tab:alpha}). Hence, our constraints are much weaker, despite being informed by a larger galaxy survey, assuming a perfect survey, and using a modelling framework that properly accounts for mass-mixing.

We strongly suspect that this discrepancy is a consequence of the fact that D23 have failed to properly account for the large halo-to-halo variance. Although their sample of $50 \times 27=1350$ merger trees used to construct their LV-SHMF may seem sufficient to beat down the halo-to-halo variance, the majority of these merger trees are not unique. This is evident from their methodology, which starts out by sampling only 280 unique merger trees, each of a different halo mass uniformly sampling the range $10^{10.5}\Msun \leq M_\rmH \leq 10^{13.3}\Msun$. However, based on Fig.~3 of D23 it is clear that the actual range of halo masses sampled by the 27 host galaxies is significantly smaller ($10^{11.7}\Msun \lta M_\rmH \lta 10^{13.0}\Msun$). Hence, based on a simple Monte-Carlo sampling test, we estimate that only $\sim 124$ of the 1350 merger trees ($\lesssim$ 10\%) in their sample are unique. This repeat-sampling of identical merger trees implies that D23 have severely underestimated the halo-to-halo variance, which has resulted in artificially narrow confidence intervals.

\section{Conclusions}
\label{sec:conclusions}

In this paper we have used a semi-analytical forward model to explore the constraining power that satellite galaxy surveys like SAGA have on the low mass end of the SHMR (roughly below a halo mass of $10^{11} \Msun$). In particular, we construct mock SAGA-like surveys by populating dark matter subhalos in the publicly available, semi-analytical \SatGen code with satellite galaxies using a flexible parametrization of the SHMR. We then analyze these mock surveys using forward models that have the statistical power of 100 SAGA-like surveys (i.e., are based on 10,000 MW-like host halos). This large number is needed to adequately beat down the host-to-host variance, which has contributions from both halo-to-halo variance at fixed halo mass and the variance in host halo masses (mass-mixing), in the model predictions. By ignoring complications due to sample incompleteness, interlopers, and observational errors, our analysis yields the most optimistic predictions for the kind of inference that is possible. Hence, the result presented below have to be interpreted as the ``best-case'' scenario. If one were to properly account for these observational effects, the constraining power will undoubtedly by weaker than what has been reported here.

Whereas D23 used the combined satellite stellar mass function, $\rmd N/\rmd M_\ast$, obtained by stacking the data of all host galaxies, in order to constrain the SHMR, we advocate instead to use (only) two numbers per host galaxy: the total number of satellites and the stellar mass (or luminosity) of the most massive (brightest) satellite. In particular, we have demonstrated that the covariance between these two observables contains valuable information regarding the scatter in host halo masses, thereby significantly reducing modeling degeneracies. Another important advantage of our methodology over simply using the combined SSMF is that it is less susciptible to systematic errors in the inferred model parameters when the prior on the unknown host halo masses is incorrect (see Fig.~\ref{fig:matrix}). The main conclusions from our parameter recovery tests are as follows:
\begin{itemize}

\item When modeling the SHMR as a simple power-law relation with constant scatter, we confirm the conclusions from earlier work \citep[][]{Garrison-Kimmel.etal.17b, Jethwa.etal.18} that have shown that the inference regarding the power-law slope, $\alpha$, is degenerate with the inferred scatter, $\sigma$. We also find that the constraints on both $\alpha$ and $\sigma$ become significantly weaker with increasing power-law slope (Fig.~\ref{fig:alpha_permutations}). 

\item The host-to-host scatter in the satellite populations of the 100 host galaxies is dominated by halo-to-halo variance in halo merger histories, and by the (unknown) scatter in host halo masses (Fig.~\ref{fig:scatter}). Only if the intrinsic scatter in stellar mass at fixed halo mass significantly exceeds 1.0dex, will the latter dominate the host-to-host scatter. However, 1~dex is much larger than the amount of scatter predicted by hydrodynamical simulations of galaxy formation and hence is unrealistic for all practical purposes. This demonstrates why it is inherently difficult to meaningfully constrain $\sigma$  using data on satellite galaxies. 

\item One of the main results from our forecasting is that there is little hope of meaningfully constraining the scatter in the SHMR, let alone any potential mass dependence thereof. Even under the most ideal circumstances of no interlopers, 100\% completeness, no observational errors, a shallow intrinsic power-law slope of $\alpha=1.5$, and a modest amount of mass mixing with a scatter of 0.15dex among the 100 host halos in the mock survey, the scatter can not be constrained to better than 0.5dex at 95\% confidence level (see Fig.~\ref{fig:sigma_permutations}). Given that the scatter at the massive end of the SHMR is constrained to be $0.15-0.20$dex, such constraints are not particularly useful. When allowing for more freedom in the SHMR, such as curvature or redshift dependence, the constraints on $\sigma$ rapidly deteriorate (see Fig.~\ref{fig:extra_freedom}). The disheartening lack of constraining power is best illlustrated by Fig.~\ref{fig:model_illustration} which demonstrates that SHMR that are intrinsically extremely different can look indistinguishable in observational parameter space.  

\item Even though SAGA comprises data on the satellite populations of 100 MW-like host galaxies, our results indicate that the sample still suffers from significant sample variance (see Fig.~\ref{fig:correlation}). This is yet another manifestation of the huge host-to-host variance to be expected for MW-like host galaxies. As a consequence, moving forward there is more merit (in terms of constraining the SHMR) in increasing the sample size, as opposed to trying to probe down to fainter, less massive satellites (see Fig.~\ref{fig:survey_limits}). We emphasize, though, that it is key to try and quantify the distribution of host halo masses as accurately as possible. 

\end{itemize}


\section*{Acknowledgments}

We are grateful to Shany Danieli, Marla Geha, Yao-Yuan Mao, and Ethan Nadler for valuable discussion, and to the anonymous referee for insightful comments that have significantly improved the manuscript. FvdB is supported by the National Science Foundation (NSF) through grant AST-2307280. This work was performed in part at the Kavli Institute for Theoretical Physics (KITP) in Santa Barbara, which is supported in part by the National Science Foundation under Grant No. NSF PHY-174895.

This work utilised, primarily for plotting purposes, the following python packages: \texttt{Matplotlib} \citep{Matplotlib_Hunter2007}, \texttt{SciPy} \citep{SciPy_Virtanen2020}, \texttt{NumPy} \citep{numpy_vanderWalt2011}, and \texttt{PyGTC} \citep{PyGTC_Bocquet2016}. 


\bibliography{references}


\appendix

\section{Summary of Results}
\label{app:results}

The tables below list the input and recovered values for the SHMR parameters for the various models discussed in the text. Each table corresponds to a specific figure in the text, as indicated. For the recovered values we list the median value plus the $5-95$ percentile range from the posterior distribution.

\begin{table}[ht]
    \vskip0.1in
    \small\centering
    \begin{tabular}{c|cc|cc}
    \hline
    \hline
      & \multicolumn{2}{c|}{Input parameters} &  \multicolumn{2}{c}{Recovered parameters}\\
    \hline
    sample & $\alpha$ & $\sigma$ &  $\alpha$ &  $\sigma$ \\
    \hline
    $S_0$ & 1.5 & 0.2 & $1.49_{-0.012}^{+0.017}$ & $0.19_{-0.124}^{+0.182}$ \\
    $S_0$ & 2.0 & 0.2 & $2.01_{-0.031}^{+0.059}$ & $0.20_{-0.179}^{+0.305}$ \\
    $S_0$ & 2.5 & 0.2 & $2.48_{-0.064}^{+0.107}$ & $0.24_{-0.211}^{+0.419}$ \\
    $S_0$ & 3.0 & 0.2 & $3.05_{-0.149}^{+0.300}$ & $0.39_{-0.354}^{+0.696}$ \\
    $S_{15}$ & 1.5 & 0.2 & $1.51_{-0.023}^{+0.029}$ & $0.25_{-0.220}^{+0.247}$ \\
    $S_{15}$ & 2.0 & 0.2 & $2.01_{-0.055}^{+0.110}$ & $0.33_{-0.295}^{+0.405}$ \\
    $S_{15}$ & 2.5 & 0.2 & $2.49_{-0.104}^{+0.125}$ & $0.32_{-0.282}^{+0.430}$ \\
    $S_{15}$ & 3.0 & 0.2 & $3.08_{-0.206}^{+0.237}$ & $0.30_{-0.272}^{+0.522}$ \\
    $S_{30}$ & 1.5 & 0.2 & $1.53_{-0.043}^{+0.046}$ & $0.23_{-0.206}^{+0.302}$ \\
    $S_{30}$ & 2.0 & 0.2 & $2.01_{-0.065}^{+0.074}$ & $0.16_{-0.141}^{+0.286}$ \\
    $S_{30}$ & 2.5 & 0.2 & $2.50_{-0.105}^{+0.132}$ & $0.19_{-0.173}^{+0.332}$ \\
    $S_{30}$ & 3.0 & 0.2 & $2.88_{-0.199}^{+0.256}$ & $0.29_{-0.256}^{+0.422}$ \\
    \hline
\end{tabular}
\caption{Results corresponding to the posteriors shown in Fig.~\ref{fig:alpha_permutations}}
\label{tab:alpha}
\end{table}

\begin{table}[ht]
    \vskip0.1in
    \small\centering
    \begin{tabular}{c|cc|cc}
    \hline
    \hline
      & \multicolumn{2}{c|}{Input parameters} &  \multicolumn{2}{c}{Recovered parameters}\\
    \hline
    sample & $\alpha$ & $\sigma$ &  $\alpha$ &  $\sigma$ \\
    \hline
    $S_0$ & 2.0 & 0.0 & $2.00_{-0.031}^{+0.054}$ & $0.21_{-0.185}^{+0.265}$ \\
    $S_0$ & 2.0 & 0.2 & $2.01_{-0.031}^{+0.059}$ & $0.20_{-0.179}^{+0.305}$ \\
    $S_0$ & 2.0 & 0.4 & $2.00_{-0.033}^{+0.082}$ & $0.25_{-0.228}^{+0.355}$ \\
    $S_0$ & 2.0 & 0.6 & $2.05_{-0.113}^{+0.096}$ & $0.75_{-0.383}^{+0.253}$ \\
    $S_{15}$ & 2.0 & 0.0 & $2.04_{-0.053}^{+0.102}$ & $0.34_{-0.304}^{+0.367}$ \\
    $S_{15}$ & 2.0 & 0.2 & $2.01_{-0.055}^{+0.110}$ & $0.33_{-0.295}^{+0.405}$ \\
    $S_{15}$ & 2.0 & 0.4 & $2.04_{-0.086}^{+0.117}$ & $0.56_{-0.462}^{+0.330}$ \\
    $S_{15}$ & 2.0 & 0.6 & $1.93_{-0.048}^{+0.086}$ & $0.33_{-0.307}^{+0.349}$ \\
    $S_{30}$ & 2.0 & 0.0 & $1.96_{-0.066}^{+0.074}$ & $0.18_{-0.168}^{+0.312}$ \\
    $S_{30}$ & 2.0 & 0.2 & $2.01_{-0.065}^{+0.074}$ & $0.16_{-0.141}^{+0.286}$ \\
    $S_{30}$ & 2.0 & 0.4 & $1.94_{-0.064}^{+0.102}$ & $0.36_{-0.330}^{+0.376}$ \\
    $S_{30}$ & 2.0 & 0.6 & $1.93_{-0.066}^{+0.107}$ & $0.35_{-0.279}^{+0.375}$ \\
    \hline
\end{tabular}
\caption{This corresponds to the ``$\sigma$ permutation" posteriors shown in Fig. \ref{fig:sigma_permutations}}
\label{tab:sigma}
\end{table}

\begin{table}[ht]
    \vskip0.1in
    \small\centering
    \begin{tabular}{c|c|ccc|ccc}
    \hline
    \hline
      & & \multicolumn{3}{c|}{Input parameters} &  \multicolumn{3}{c}{Recovered parameters}\\
    \hline
    model & DoF & $\alpha$ & $\sigma$ & free parameter & $\alpha$ & $\sigma$ & free parameter \\
    \hline
    fiducial & 2 & 2.0 & 0.2 & - & $2.01_{-0.054}^{+0.106}$ & $0.33_{-0.292}^{+0.399}$ & - \\
    $\beta$ free & 3 & 2.0 & 0.2 & 0.0 & $2.59_{-0.738}^{+1.269}$ & $0.84_{-0.771}^{+0.560}$ & $0.20_{-0.266}^{+0.410}$ \\
    $\gamma$ free & 3 & 2.0 & 0.2 & 0.0 & $2.17_{-0.329}^{+0.591}$ & $0.56_{-0.504}^{+0.547}$ & $0.16_{-0.405}^{+0.365}$ \\
    $\nu$ free & 3 & 2.0 & 0.2 & 0.0 & $2.12_{-0.132}^{+0.238}$ & $0.24_{-0.215}^{+0.374}$ & $-0.16_{-0.276}^{+0.148}$ \\
    \hline
\end{tabular}
\caption{Results corresponding to the posteriors shown in Fig.~\ref{fig:extra_freedom}}
\label{tab:free}
\end{table}

\begin{table}[ht]
    \vskip0.1in
    \small\centering
    \begin{tabular}{c|c|cc|cc}
    \hline
    \hline
      & & \multicolumn{2}{c|}{Input parameters} &  \multicolumn{2}{c}{Recovered parameters}\\
    \hline
    $\mlim$ & $\Nhost$ & $\alpha$ & $\sigma$ &  $\alpha$ &  $\sigma$ \\
    \hline
    5.5 & 100 & 2.0 & 0.2 & $2.01_{-0.040}^{+0.066}$ & $0.35_{-0.300}^{+0.396}$ \\
    6.5 & 100 & 2.0 & 0.2 & $2.01_{-0.055}^{+0.110}$ & $0.33_{-0.295}^{+0.405}$ \\
    7.5 & 100 & 2.0 & 0.2 & $2.04_{-0.083}^{+0.128}$ & $0.19_{-0.173}^{+0.366}$ \\
    6.5 & 300 & 2.0 & 0.2 & $2.01_{-0.026}^{+0.040}$ & $0.19_{-0.158}^{+0.240}$ \\
    6.5 & 10000 & 2.0 & 0.2 & $2.01_{-0.015}^{+0.021}$ & $0.16_{-0.128}^{+0.155}$ \\
    \hline
\end{tabular}
\caption{Results corresponding to the posteriors shown in Fig.~\ref{fig:survey_limits}}
\label{tab:future}
\end{table}

\begin{table}[ht]
    \vskip0.1in
    \small\centering
    \begin{tabular}{c|cc|cc|cc}
    \hline
    \hline
      & \multicolumn{2}{c|}{Input parameters} &  \multicolumn{2}{c|}{Our Framework} & \multicolumn{2}{c}{D23 Framework}\\
    \hline
    data $\times$ model & $\alpha$ & $\sigma$ & $\alpha$ & $\sigma$ & $\alpha$ & $\sigma$ \\
    \hline
    $S_{0} \times S_{0}$ & 2.0 & 0.2 & $1.99_{-0.024}^{+0.048}$ & $0.14_{-0.128}^{+0.288}$ & $2.00_{-0.035}^{+0.049}$ & $0.25_{-0.145}^{+0.236}$ \\
    $S_{0} \times S_{15}$ & 2.0 & 0.2 & $2.02_{-0.041}^{+0.048}$ & $0.15_{-0.133}^{+0.262}$ & $2.02_{-0.031}^{+0.041}$ & $0.16_{-0.142}^{+0.257}$ \\
    $S_{0} \times S_{30}$ & 2.0 & 0.2 & $2.01_{-0.054}^{+0.069}$ & $0.14_{-0.129}^{+0.279}$ & $2.11_{-0.032}^{+0.040}$ & $0.14_{-0.127}^{+0.231}$ \\
    $S_{15} \times S_{0}$ & 2.0 & 0.2 & $1.96_{-0.020}^{+0.102}$ & $0.08_{-0.077}^{+0.561}$ & $2.00_{-0.035}^{+0.057}$ & $0.34_{-0.149}^{+0.223}$ \\
    $S_{15} \times S_{15}$ & 2.0 & 0.2 & $2.02_{-0.044}^{+0.066}$ & $0.22_{-0.191}^{+0.322}$ & $2.01_{-0.032}^{+0.049}$ & $0.20_{-0.178}^{+0.272}$ \\
    $S_{15} \times S_{30}$ & 2.0 & 0.2 & $2.04_{-0.063}^{+0.076}$ & $0.18_{-0.163}^{+0.310}$ & $2.09_{-0.029}^{+0.039}$ & $0.12_{-0.106}^{+0.260}$ \\
    $S_{30} \times S_{0}$ & 2.0 & 0.2 & - & - & $1.92_{-0.063}^{+0.070}$ & $0.59_{-0.260}^{+0.216}$ \\
    $S_{30} \times S_{15}$ & 2.0 & 0.2 & $1.91_{-0.050}^{+0.067}$ & $0.64_{-0.189}^{+0.215}$ & $1.91_{-0.055}^{+0.082}$ & $0.46_{-0.334}^{+0.276}$ \\
    $S_{30} \times S_{30}$ & 2.0 & 0.2 & $1.97_{-0.069}^{+0.084}$ & $0.28_{-0.246}^{+0.393}$ & $1.94_{-0.029}^{+0.058}$ & $0.21_{-0.186}^{+0.308}$ \\
    \hline
\end{tabular}
\caption{Results corresponding to the $3 \times 3$ set of posteriors shown in Fig.~\ref{fig:matrix}}
\label{tab:matrix}
\end{table}

\section{Mass Accretion Histories}
\label{app:mah}

To illustrate how variance in mass accretion histories impacts our two summary statistics, $\nsat$ and $\mstarmax$, we proceed as follows. We populate all $\Nhost=10^4$ host halos in each of our three samples, $S_0$, $S_{15}$ and $S_{30}$, with satellite galaxies, using a deterministic version of our fiducial model with $\alpha=2.0$ and $\sigma=0.0$; hence, there is no variance in $\nsat$ and $\mstarmax$ due to scatter in the SHMR. For each host halo we also use its merger tree to determine the redshift, $z_{50}$, at which the mass of its main progenitor first reaches 50\% of its final ($z=0$) mass.
\begin{figure*}
    \centering
    \includegraphics{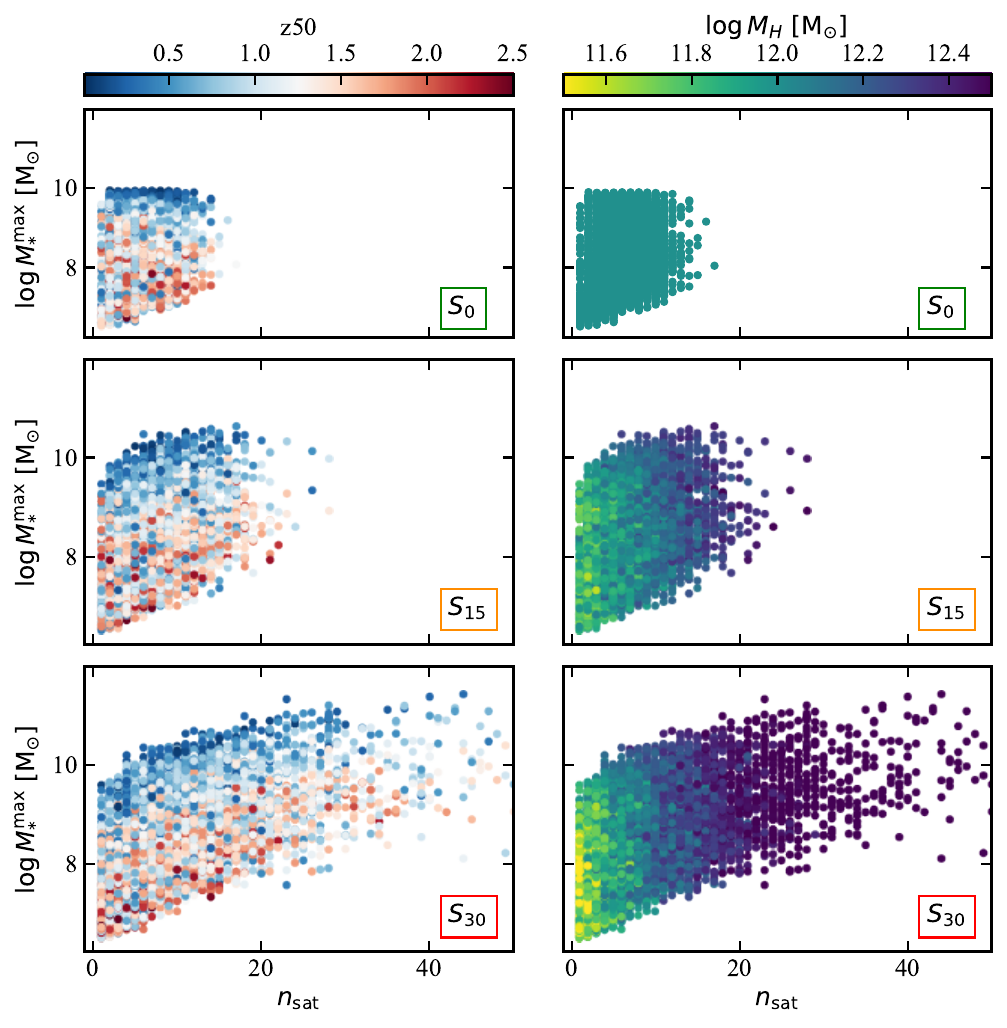}
    \caption{The $\nsat$- $\mstarmax$ distributions for the complete sets of $\Nhost=10^4$ merger tree realizations for samples $S_0$, $S_{15}$, and $S_{30}$, constructed using a deterministic version of our fiducial SHMR (i.e., without scatter). The left-hand column of panels is colored by the half-mass assembly redshift ($z_{50}$), where ``late forming" halos are blue and ``early forming" are red. Similarly, the right-hand column of panels is colored by host halo mass ($M_H$), where yellow halos are less massive and purple halos are more massive. Note how scatter in half-mass assembly time predominantly contributes variance in $\mstarmax$, while variance in $\nsat$ mainly originates from 
    scatter in host halo mass.}
    \label{fig:mah_variance}
\end{figure*}

Fig.~\ref{fig:mah_variance} shows scatter plots of the resulting $\nsat$ and $\mstarmax$ color-coded either by the half-mass assembly redshift ($z_{50}$, left-hand panels) or the present day mass ($M_\rmH$, right-hand panels) of the host halo. As is evident, variance in $n_{\rm sat}$ and $\mstarmax$ are predominantly due to scatter in $M_\rmH$ and $z_{50}$, respectively. These trends are easy to understand. To good approximation, the number of subhalos with an accretion mass, $M_{\rm acc}$, above a given value scales linearly with host halo mass \citep[e.g.,][]{Jiang.vdBosch.16}. Hence, for a given SHMR, more massive host halos will have more satellites above a given stellar mass limit. The trend with halo assembly time arises because halos that have recently gained much of their mass (``later-forming" halos) are more likely to have done so through a major merger, i.e., by accreting a relatively massive subhalo which, for a given SHMR, will host a relatively massive satellite.

Note also that including more host halo mass mixing (from top to bottom), skews the $\nsat$- $\mstarmax$ distributions further up and to right, which causes an increase in the Pearson rank-order correlation coefficient, $r_\rmP$ (cf. Fig.~\ref{fig:correlation}). This illustrates how including information on this correlation is informative for constraining the underlying distribution in host halo masses (see discussion in Section~\ref{sec:mass_mixing}).

\section{Choice of summary statistics}
\label{app:stats}

As discussed in the main text, our method for modelling a SAGA-like survey only uses the total number of satellites, $\nsat$, and the stellar mass of the most massive satellite, $\mstarmax$. This ignores additional information regarding the stellar masses of the other satellites, which in principle could help to tighten the constraints on the SHMR parameters.
\begin{figure*}
    \centering
    \includegraphics{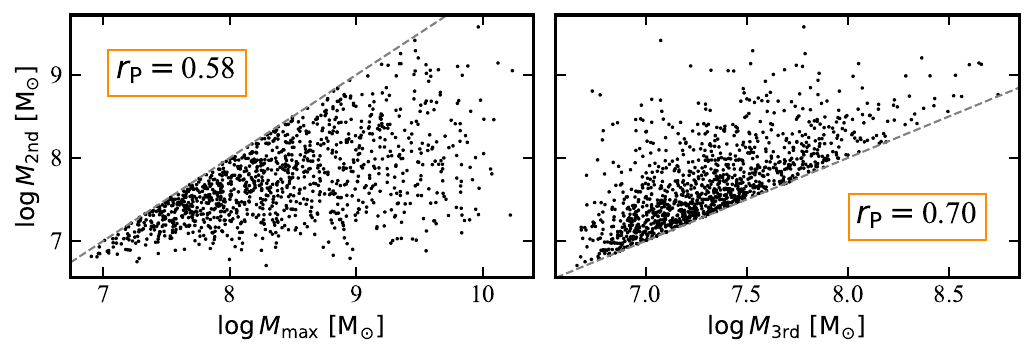}
    \caption{Scatter plot of subsequent mass ranks for the $S_{15}$ sample constructed using the deterministic version of our fiducial SHMR parametrization ($\alpha=2.0, \sigma=0.0$) and conditioned on $\nsat = 6$. The grey dashed line shows the one-to-one correspondence. The left column shows the most massive satellite against the second most massive satellite per host. Similarly, the right column shows the second most massive satellite against the third most massive satellite per host. The corresponding Pearson’s rank-order correlation coefficients are indicated. As expected, these observables are strongly correlated, which has to be accounted for when folding in these data. See text for discussion.}
    \label{fig:mass_rank_correlation}
\end{figure*}

An obvious extension of our method therefore would be to fold in information on the mass of the second most massive satellite per host, $M_{*}^{2{\rm nd}}$ and/or the $n^{\rm th}$ most massive satellite per host, $M_{*}^{n{\rm th}}$. However, including such information in our likelihood estimation is far from trivial, mainly because the stellar masses of the mass-ranked satellites are strongly correlated. This is illustrated in Fig.~\ref{fig:mass_rank_correlation} which plots  $M_{*}^{2 {\rm nd}}$ vs. $\mstarmax$ (left-hand panel) and $M_{*}^{2 {\rm nd}}$ vs. $M_{*}^{3 {\rm rd}}$ (right-hand panel) in a mock data set consisting of $10^4$ equal mass host haloes taken from sample $S_0$ and constructed using the same deterministic version of our fiducial SHMR ($\alpha=2.0, \sigma=0.0$) as used in Appendix~\ref{app:mah}. Recall that we evaluate the likelihood of $\mstarmax$ at bins of fixed $\nsat$, which is why we show the results for the subset of hosts with $\nsat=6$.  This particular value for $\nsat$ is arbitrary, and results for other $\nsat \geq 3$ bins are very similar.  As can be seen, and is evident from the large Pearson rank-order correlation coefficients, the rank-ordered stellar masses are strongly correlated. In order to properly take this into account, the likelihood function needs to have the form
\begin{equation}\label{lnLsum}
    \ln\calL(\bD|\btheta) = \ln\calL_N(\bD|\btheta)+ \ln\calL_{M_{*}^{\rm max}|N}(\bD|\btheta)
    + \ln\calL_{M_{*}^{2{\rm nd}}| M_{*}^{\rm max}, N} (\bD|\btheta)
    + \ln\calL_{M_{*}^{3{\rm rd}}| M_{*}^{2{\rm nd}}, M_{*}^{\rm max}, N} (\bD|\btheta)\,
\end{equation}
i.e., the likelihood for the $n^{\rm th}$ most massive satellite is conditional on the stellar masses of the satellites of rank-order $<n$.

The main issue is that the number of host haloes used to compute the model predictions needs to be sufficiently large such that the model uncertainty is small compared to the uncertainty in the data. As we demonstrate in Appendix~\ref{app:convergence}, our two summary statistics ($\nsat$ and $\mstarmax$) already require $\hat{N}_{\rm host} \gta 10^4$. Therefore by including additional data, the minimum value of $\hat{N}_{\rm host}$ that is required rapidly becomes too large for practical purposes. This is because the computation of a conditional likelihood like $\ln\calL_{M_{*}^{3{\rm rd}} | M_{*}^{2{\rm nd}}, M_{*}^{\rm max}, N} (\bD|\btheta)$ requires binning the data (and the model) in bins of $\nsat$, $M_{*}^{2{\rm nd}}$ and $M_{*}^{\rm max}$. Assuring that there are a sufficient number of systems in each of these bins to make a sufficiently accurate prediction for the model requires large $\hat{N}_{\rm host}$. Very roughly, for every rank-ordered stellar mass added to the likelihood, $\hat{N}_{\rm host}$ has to increase by an order of magnitude. Hence, going to rank-order three, as in equation~(\ref{lnLsum}) requires $\hat{N}_{\rm host} \gta 10^6$. Given that the CPU time required for the analysis scales linearly with $N_{\rm host}$, this becomes unfeasible. An alternative is to use unconditional likelihoods, but this requires computing an accurate covariance matrix for the model, which faces similar challenges. This explains why our method only uses $\nsat$ and $\mstarmax$ in order to constrain the SHMR.
 
\begin{figure*}
    \centering
    \includegraphics{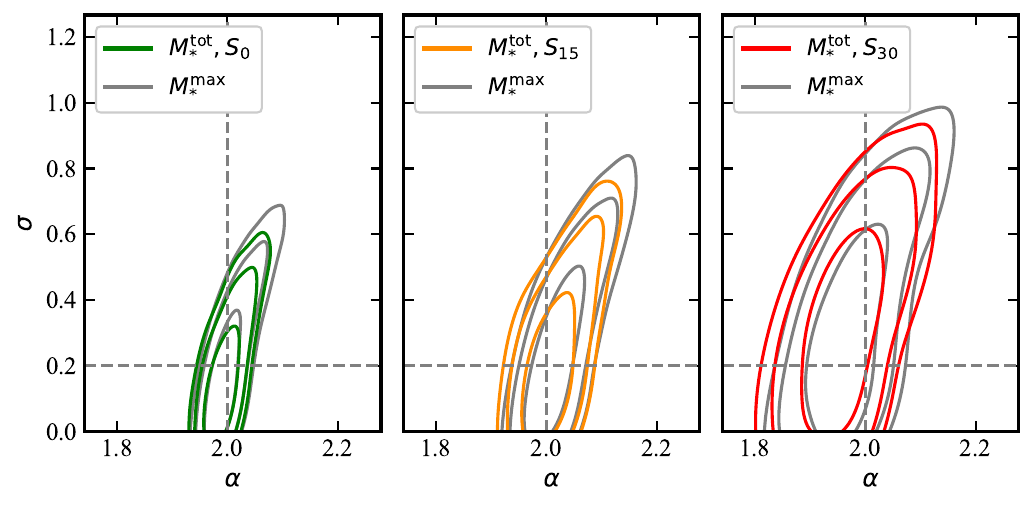}
    \caption{Posterior constraints on $\alpha$ and $\sigma$ obtained for our fiducial mock data sets for samples $S_0$ (left panel), $S_{15}$ (middle panel) and $S_{30}$ (right panel). Grey-colored contours show the 68, 95 and 99\% confidence level obtained using our standard statistical framework based on $\nsat$ and $\mstarmax$ (same as the blue ``self-consistent" contours along the diagonal in Fig.~\ref{fig:matrix}). For comparison, the colored contours show the corresponding constraints obtained when replacing $\mstarmax$ with the total stellar mass of all $\nsat$ satellites, $M_{\ast}^{\rm tot}$. Note the similarity of the posterior constraints in the two cases. See text for discussion.}
    \label{fig:added_stats}
\end{figure*}

To gauge the robustness of our standard method, we have experimented with replacing $\mstarmax$ with the \textit{total} mass of all $\nsat$ satellites, $M_{\ast}^{\rm tot}$. The results for each of our three fiducial mock data samples are shown in Fig.~\ref{fig:added_stats}, which compares the posteriors obtained using our standard method using $\mstarmax$ in grey to the posteriors obtained when using the total stellar mass instead (in color). As is evident, the results are virtually indistinguishable. Hence, both $\mstarmax$ and $M_{\ast}^{\rm tot}$ have a similar constraining power. As argued in the main text (see Section~\ref{sec:stats}), this (i) indicates that adding additional information on $M_{*}^{\rm 2nd}$, $M_{*}^{\rm 3rd}$, etc. is unlikely to significantly improve the posterior constraints, and (ii) is likely due to the self-similar, universal shape of the subhalo mass function.

\section{How many merger trees does one need?}
\label{app:convergence}

Throughout the main text we have used $\hat{N}_{\rm host}=10^4$ merger trees for each of our model evaluations, arguing that this is required in order to make reliable inference regarding the SHMR parameters. Recall (see Section~\ref{sec:stats}) that the likelihood for a host to have $\nsat$ satellites is inferred form the frequency distribution of that number of satellites among the $\hat{N}_{\rm host}$ host realizations. If none of the $\hat{N}_{\rm host}$ hosts have that number of satellites the inferred likelihood is zero and the model is rejected. Now suppose the model has the same SHMR parameters, $\btheta$ as the data, but $\hat{N}_{\rm host}$ is very small. It can happen that the model is incorrectly rejected simply because none of the $\hat{N}_{\rm host}$ merger trees used for the model prediction have the same $\nsat$ as one (or more) of the hosts in the data sample. It is prudent therefore, that $\hat{N}_{\rm host}$ is sufficiently large that such an incorrect rejection of the correct model becomes negligibly small.

In order to assess what value of $\hat{N}_{\rm host}$ is required, we proceed as follows. To start, for each sample ($S_0$, $S_{15}$ and $S_{30}$), all $10,100$ merger trees are pushed through our fiducial parametrization of the SHMR ($\alpha=2.0, \sigma=0.2$), which we use to create 101 unique mock SAGA surveys per sample. For a given mock data set, a likelihood value can be measured using the remaining trees as the model. In our standard method, we use {\it all} the remaining $\hat{N}_{\rm host} = 10^4$ merger trees to do so. However, here we  vary $\hat{N}_{\rm host}$ in order to test how that impacts the erroneous model rejection mentioned above. Note that when varying $\hat{N}_{\rm host}$, we can only measure $10^4/\hat{N}_{\rm host}$ different likelihoods per mock data set. Next, we define $f_{\rm rej}$ as the fraction of those likelihoods for which the likelihood is zero (i.e., for which the correct model would be erroneously rejected). Since we have $101$ different mock data sets, we can repeat this exercise 101 times, thus obtaining a distribution of $f_{\rm rej}$ values from which we can infer the (mock) data-to-data scatter.
\begin{figure*}
    \centering
    \includegraphics{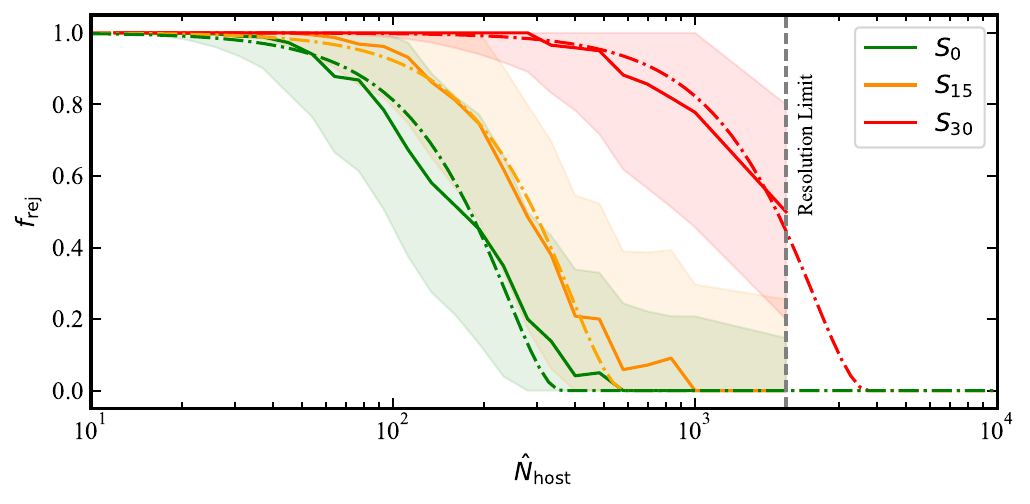}
    \caption{The fraction of models, $f_{\rm rej}$, that is erroneously rejected due to an insufficient sampling of merger trees as a function of the number of host halo merger trees, $\hat{N}_{\rm host}$, used. Solid lines show the median values of $f_{\rm rej}$ obtained from 101 different mock data sets, with different colors corresponding to different mock samples, as indicated. The shaded regions indicate the corresponding $16-84$ percentile ranges. The dash-dotted lines show smoothed step function fits to the $f_{\rm rej}$ data, while the grey vertical dashed line shows the limit we impose to account for small number statistics. As discussed in the text, this demonstrates that our methodology requires $\hat{N}_{\rm host}=10^4$ in order to make a reliable inference about the SHMR given a SAGA-like data sample.}
    \label{fig:convergence}
\end{figure*}

The solid, colored lines in Fig.~\ref{fig:convergence} plot the resulting $f_{\rm rej}$ as a function of $\hat{N}_{\rm host}$ for the three samples, as indicated. The shaded regions indicate the 16-84 percentile ranges obtained from the 101 different data samples. Note that as $\hat{N}_{\rm host}$ increases, the number of independent likelihood evaluations per data sample decreases, which is why we don't show any results past $\hat{N}_{\rm host} = 2000$ (indicated by the dashed grey vertical line, where for each mock data sample we have to estimate $f_{\rm rej}$ from only 5 likelihood evaluations). As expected, $f_{\rm rej}$ decreases with $\hat{N}_{\rm host}$. When $\hat{N}_{\rm host} \lta 50$ the correct model for the SHMR is almost guaranteed to be erroneously rejected; in this case the posterior constraints will vary drastically from one mock data set to the next. In other words, any results derived from these models would be utterly unreliable. Based on a rough extrapolation (shown by the dash-dotted lines), and accounting for the data-to-data scatter, we estimate that having $f_{\rm rej} \lta 0.01$ requires $\hat{N}_{\rm host} \gta 10^4$. We emphasize that these conclusions are specific to the method of forward modeling presented here. Different methods to constrain the SHMR may require different $\hat{N}_{\rm host}$.

\end{document}